\documentclass[useAMS,usenatbib]{mn2e}
\usepackage{epsfig, color}

\title[Pop III Binaries]
      {Constraining the Statistics of Population III Binaries}

\author[A. Stacy and V. Bromm]
       {Athena Stacy$^{1}$\thanks{E-mail: athena.stacy@nasa.gov} and Volker Bromm$^{2}$\\
        $^{1}$NASA Goddard Space Flight Center, Greenbelt, MD 20771, USA \\
        $^{2}$Department of Astronomy and Texas Cosmology Center, University of Texas, Austin, TX 78712, USA \\}

\pagerange{\pageref{firstpage}--\pageref{lastpage}}
\pubyear{2012}

\begin{document}

\maketitle
\topmargin-1cm

\label{firstpage}

\begin{abstract}
We perform a cosmological simulation in order to model the growth and evolution of Population III (Pop III) stellar systems in a range of host minihalo environments.  A Pop III multiple system forms in each of the ten minihaloes, and the overall mass function is top-heavy compared to the currently observed initial mass function in the Milky Way.  Using a sink particle to represent each growing protostar, we examine the binary characteristics of the multiple systems, resolving orbits on scales as small as 20 AU.  We find a binary fraction of $\sim$ 35\%,  with semi-major axes as large as 3000 AU. The distribution of 
orbital periods is slightly peaked at $\la900$ yr,
while the distribution of mass ratios is relatively flat.  Of all sink particles formed within the ten minihaloes, $\sim$ 50\% are lost to mergers with larger sinks, and $\sim$ 50\% of the remaining sinks are ejected from their star-forming disks. The large binary fraction may have important implications for Pop~III evolution and nucleosynthesis, as well as the final fate of the first stars.     
\end{abstract}

\begin{keywords}
stars: formation - Population III - galaxies: formation - cosmology: theory - first stars - early Universe 
\end{keywords}

\section{Introduction}
The period during which the first stars arose marked a pivotal turning point in the evolution of the Universe.  The first stars, also known as Population III (Pop III), are believed to have formed at $z\ga20$ within dark matter (DM) minihaloes of 10$^6$ M$_{\odot}$
 (e.g. \citealt{haimanetal1996,tegmarketal1997,yahs2003}).  
Afterwards they generated ionizing photons that contributed to the reionization of the Universe
(e.g. \citealt{kitayamaetal2004,syahs2004,whalenetal2004,alvarezetal2006,johnsongreif&bromm2007}),
while those Pop III stars that ended their lives as supernovae (SNe) contributed to the metallicity of the intergalactic medium
 (IGM; \citealt{madauferrara&rees2001,moriferrara&madau2002,brommyoshida&hernquist2003,wada&venkatesan2003,normanetal2004,tfs2007,greifetal2007,greifetal2010,wise&abel2008,maioetal2011}; recently reviewed in \citealt{karlssonetal2013}).  This metal-enrichment and growing radiation background provided by the first stars determined the environment in which later stellar generations formed.  



The mass of a given Pop III star plays the central role in determining how the star will influence its surroundings.  
For instance, more massive stars are more efficient in producing ionizing photons, per unit stellar mass, than low-mass stars (\citealt{brommetal2001,schaerer2002}). 
Furthermore, non-rotating primordial stars with mass 140~M$_{\odot}$~$<$~$M_{*}$~$<$~260~M$_{\odot}$ are believed to have exploded as pair-instability supernovae (PISNe; \citealt{heger&woosley2002}),
releasing the entirety of their metal content into the IGM and surrounding haloes, while  stars within the range  15~M$_{\odot}$~$<$~$M_{*}$~$<$~40 M$_{\odot}$ ended their lives as core-collapse SNe.  
On the other hand, non-rotating Pop III stars with main sequence masses in the range 40~M$_{\odot}$~$<$~$M_{*}$~$<$~140~M$_{\odot}$ or  $M_{*}$~$>$~260~M$_{\odot}$ are expected to collapse directly into black holes (BHs), thus contributing no metals to their surroundings.

BH remnants of Pop III stars may emit copious amounts of radiation during subsequent mass accretion, however.  
While this radiation will heat and ionize its local surroundings, the X-ray component has a long mean-free path that allows it to ionize the IGM and minihalo gas as far as $10-100$ kpc away.  
This contributed to reionization, while the boosted electron fraction can catalyze the formation of additional H$_2$ molecules, mildly enhancing the global cooling and star-formation rate 
(e.g. \citealt{madauetal2004,ricotti&ostriker2004,kuhlen&madau2005,haiman2011,jeonetal2012}).  
Recent work has demonstrated that a BH remnant which has a stellar binary companion, giving rise to a high-mass X-ray binary (HMXB), will exert stronger feedback on its surroundings than 
an isolated BH (e.g. \citealt{mirabeletal2011,jeonetal2012}).
This is because accretion onto solitary BHs is more susceptible to feedback that in turn reduces the BHs accretion rate and luminosity (e.g. \citealt{johnson&bromm2007,alvarezetal2009,milosetal2009}), whereas the near-Eddington luminosity of a HMXB can persist through Roche lobe overflow.
In this way the binary characteristics of Pop III stars will influence the nature of Pop III feedback.   
Also intriguing are observations indicating that ultra-luminous X-ray sources (ULXs), which may be accreting, intermediate-mass BHs, occur at a higher rate in less massive and lower-metallicity galaxies (e.g. \citealt{swartzetal2008}).  Recent modeling of binary system evolution over a range of metallicity also supports a higher rate of ULX formation in lower-metallicity star clusters (\citealt{lindenetal2010}). 
We note that BH binaries are detectable in channels other than electromagnetic radiation. For instance, \cite{kowalskaetal2012} find that if Pop III stars left behind a significant BH population, then their gravitational wave (GW) emission will dominate the low-frequency ($< 100$Hz) spectrum as long as the Pop III binary fraction was at least $10^{-2}$.

A binary companion can also influence a star's evolution during and after its main sequence (MS) stage, as well as the type of death the star will undergo 
(e.g. \citealt{wellsteinetal2001,petrovicetal2005}, \nocite{petrovicetal2005b}2005b; \citealt{cantielloetal2007,deminketal2009,langer2012,sanaetal2012}).  
For instance, massive stars that leave behind BH remnants may power collapsar gamma-ray bursts (GRBs; e.g. \citealt{woosley1993,lee&ramirez2006}).
However, this requires sufficient
angular momentum in the Pop~III progenitor for an accretion torus to
form around the remnant BH.   The progenitor star
must also lose its hydrogen envelope to enable the relativistic jet to
penetrate through and exit the star (e.g. \citealt{zhangetal2004}, but see \citealt{suwa&ioka2011,nakauchietal2012}).
One way to fulfill both of these conditions is for the GRB progenitor to have a close binary companion that removes the hydrogen envelope due to the heating during a common-envelope
phase (e.g. \citealt{leeetal2002,izzardetal2004}). 


A star which receives mass from or merges with its binary companion potentially undergoes a process of `rejuvenation' in which its central hydrogen abundance increases  (e.g. \citealt{hellings1983, braun&langer1995,dray&tout2007}).  This causes the star's apparent age to be less than its true age, and may lead the star to appear as a blue straggler on the Hertzsprung-Russel diagram (\citealt{langer2012}). 
Whether rejuvenation will result from mass transfer depends on the evolutionary stage of the binary components, and is not expected to occur if one of the members is already post MS.  For instance, a star with an undermassive helium core may result instead (e.g. \citealt{braun&langer1995}).  Stellar mergers may furthermore lead to the generation of large-scale magnetic fields observed within massive MS stars (e.g. \citealt{donati&landstreet2009,ferrarioetal2009}).  In addition, while single stars must typically have masses $\ga$ 20 M$_{\odot}$ to go through a Wolf-Rayet (WR) phase (\citealt{hamann&liermann2006}), and even larger masses at lower metallicities (\citealt{meynet&maeder2005}), a member of a close binary may become a WR star at a lower mass of $\sim$ 10 M$_{\odot}$ (e.g. \citealt{vanbeverenetal1998,crowther2007}).

A star that gains mass from its binary companion through an accretion disk may spin up to critical rotation speed, and rapid rotation can alter a star's evolution in a number of ways 
(see e.g. \nocite{stacyetal2011}Stacy et al. 2011a; \citealt{stacyetal2013,maeder&meynet2012,yoonetal2012}).
Rotation will allow for rotationally induced mixing and possibly chemically homogeneous evolution (CHE). 
CHE, in turn, may lower the minimum mass at which a star will undergo a PISN to as low as $\sim$ 65 M$_{\odot}$ (e.g. \citealt{chatz&wheeler2012}).  
If a rotating star instead leaves behind a BH remnant, a collapsar GRB may result even without a binary companion.
This is because CHE may allow stars to bypass the red giant phase to become a WR
star. This evolutionary path may furthermore allow the star to retain
enough angular momentum to become a GRB  (e.g. \citealt{yoon&langer2005,woosley&heger2006}).
Rotationally induced chemical mixing will generally enhance a star's luminosity, temperature, 
and metal production, though this depends sensitively on the assumed magnetic field evolution within the star (e.g. \citealt{ekstrometal2008,yoonetal2012}).
Another possible consequence of both rotation and mass transfer is enhancement in surface nitrogen abundance 
(e.g. \citealt{ekstrometal2008,langeretal2008,brottetal2011,langer2012}).

Observations of massive stars in the Milky Way find that $\sim$ 45-75\% of O-stars have spectroscopic binary companions (\citealt{masonetal2009, sana&evans2011}), and calculations of binary evolution furthermore find that 20-30\% of O-stars will merge with their companions (\citealt{sanaetal2012}).  Binarity therefore plays a crucial role in the evolution of massive stars in the Milky Way.  While similar observations cannot be made for Pop III stars, we may still study them numerically.
Early computational studies of Pop~III star formation indicated that they would grow to be highly massive
($\ga 100$ M$_{\odot}$; e.g. \citealt{abeletal2002,brommetal2002,bromm&loeb2004,yoh2008}), and that each minihalo would 
host only a single Pop III star.  
More recent work, however, has found that primordial star-forming gas will undergo continued fragmentation well after the initial star has formed.
 Pop III multiplicity has been found both in small-scale simulations that utilize idealized initial conditions
 (e.g. \citealt{machidaetal2008},\nocite{clarketal2008,clarketal2011a} Clark et al 2008; 2011a),
 and in simulations that are initialized on cosmological scales
  (e.g. \citealt{turketal2009,stacyetal2010}; Clark et al. 2011b, \nocite{clarketal2011b} \citealt{greifetal2011, greifetal2012})
 and also when accounting for radiative feedback (e.g. \citealt{smithetal2011,stacyetal2012}).  
 Recent numerical advances have thus shown that the processes that lead to binary formation apply to both enriched and metal-free gas (e.g. \citealt{tohline2002}), 
 and a new picture has emerged in which Pop III stars typically form in multiples and within disks.

In order to determine the role of multiplicity in Pop~III evolution and feedback, in this study we aim to better constrain the nature of Pop III binaries.  We do this through simulating a large cosmological box which contains approximately ten minihaloes in which primordial stars form by $z\simeq 20$. 
We examine the rate of binary formation and protostellar mergers, the typical characteristics of the binaries, and the overall mass function 
at this stage of evolution of the Pop III systems.
Stars ejected from their Pop III cluster may stop accreting and remain low-mass and long-lived to the present day.  Such stars may be observed in dwarf galaxies or the Milky Way halo, so we furthermore discuss the frequency of low-mass sink ejections in our suite of minihaloes.
In Section 2 we describe our numerical methods, while we present our results in Section 3.  
We further test our results with a resolution study in Section 4.
We address our caveats in Section 5, and conclude in Section 6.

\begin{figure*}
\includegraphics[width=.8\textwidth]{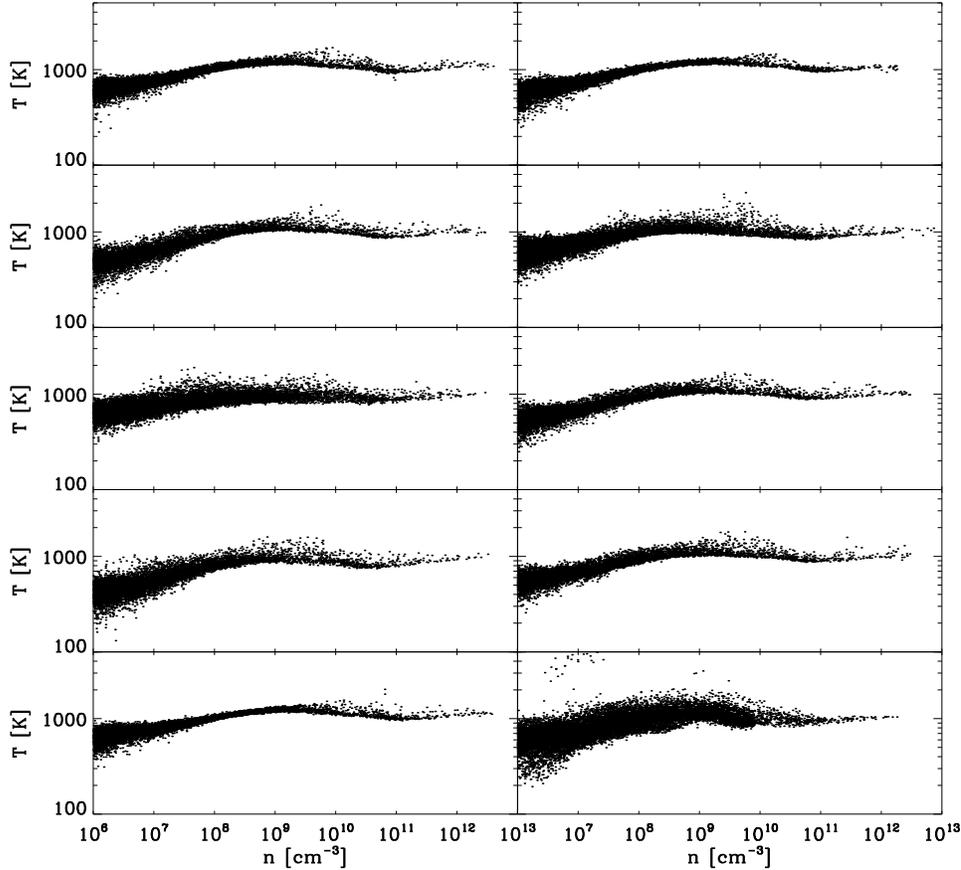}
 \caption{Temperature versus number density  of the gas just prior to the formation of the first sink particle in each minihalo.
The similarity in the early thermal evolution is striking. At $n>10^8$ cm$^{-3}$, three-body reactions rapidly convert the gas into fully molecular form. The corresponding boost in H$_2$ cooling balances adiabatic heating, resulting in nearly isothermal evolution. H$_2$ line opacity will become important only at even higher densities, eventually causing deviations  
from the near-isothermality seen here.
}
\label{Tvsn}
\end{figure*}

\begin{figure*}
\includegraphics[width=.8\textwidth]{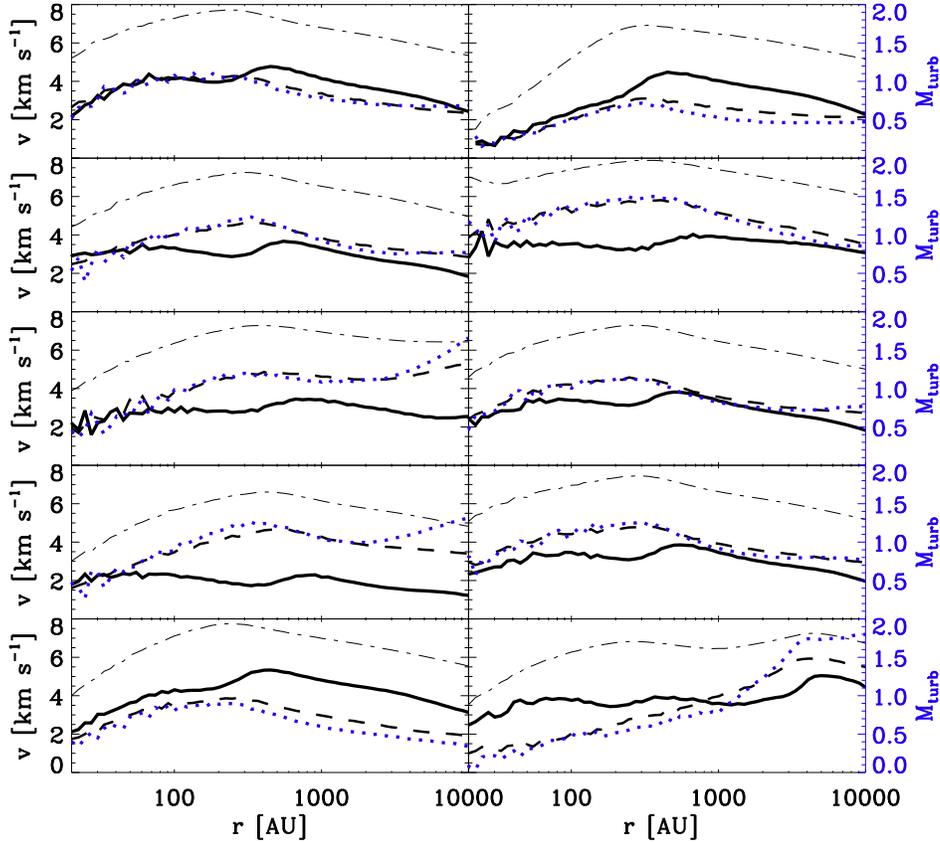}
 \caption{Velocity structure of the gas just prior to the formation of the first sink particle in each minihalo.
Minihaloes are shown in order going from top left to bottom right.
Solid lines show radial velocity $v_{\rm rad}$, dashed lines depict rotational velocity $v_{\rm rot}$, and dash-dotted lines show the free-fall velocity $v_{\rm ff}$.  
Blue dotted lines represent $M_{\rm turb}$ and follow the scaling shown on the right-hand y-axis.  
Turbulence is generally sonic ($M_{\rm turb}$=1), though in some cases the level of turbulence can become mildly supersonic.
Over these distance scales,  $v_{\rm rad}$ and $v_{\rm rot}$ both remain on the order of $\sim 1/2$ of $v_{\rm ff}$.
While the H$_2$ chemistry leads to very similar thermal histories for each minihalo, differences in mass inflow history yield a greater variety in the velocity structure of the inner gas.  
In particular, the relative magnitude of $v_{\rm rad}$ and $v_{\rm rot}$ differs in each case.  Though they remain within a factor of two of each other, $v_{\rm rad}$ dominates in some cases while $v_{\rm rot}$ dominates in others.
}
\label{mini_vel}
\end{figure*}

\section{Numerical Methodology}

\subsection{Initial Setup}
We perform our investigation using  {\sc gadget-2,} a widely-tested three-dimensional N-body and SPH code (\citealt{springel2005}). 
We begin with a 1.4~Mpc (comoving) box containing 512$^3$ SPH gas particles and the same number of DM particles.  The simulation is initialized at $z=100$.   Positions and velocities are assigned to the particles in accordance with a 
$\Lambda$CDM cosmology with $\Omega_{\Lambda}=0.7$, $\Omega_{\rm M}=0.3$, $\Omega_{\rm B}=0.04$, $\sigma_8=0.9$, and $h=0.7$.   Each gas particle has a mass of $m_{\rm sph} = 120$ M$_{\odot}$, while DM particles  have a mass of  $m_{\rm DM} = 770$ M$_{\odot}$.  Once a gas particle reaches a threshold density of $10^3$ cm$^{-3}$, 
we employ `marker sinks' instead of following these regions to increasingly higher densities.  
This allows us to efficiently evolve the entire box through a time period over which ten minihaloes form.
Though the marker sinks are numerically similar to the protostellar sink particles to be described in Section 2.4, note that the marker sinks 
are implemented at much lower densities than the protostellar sink particles of the final highest-resolution calculations.  

The location of the marker sinks coincides with the centers of the minihaloes.  
In particular, we locate the first ten sink particles which form within the first ten minihaloes.
Once these positions are ascertained, the cosmological box is reinitialized at $z=100$ for each individual minihalo, but with 64 `child' particles added around a 100-140 kpc region where the target halo will form. Larger regions of refinement are used for minihaloes whose mass originates from a larger area of the cosmological box.  Particles at progressively larger distances from the minihalo are given increasingly larger masses, such that 
in the refined initial conditions there is a total of $\sim$ 10$^7$ particles.
The most resolved particles are of mass $m_{\rm sph} = $1.85~M$_{\odot}$ and $m_{\rm DM} = 12$~M$_{\odot}$.

\subsection{Cut-Out and Refinement Technique}
The refined simulations are run until the gas reaches a density of $5 \times 10^7$ cm$^{-3}$.  All particles beyond 10 physical pc are then removed from the simulation.  Once the gas has reached these densities, this central star-forming region is gravitationally bound, and the effects of the outer minihalo and neighboring haloes can be safely ignored.
Also note that the gas at the cut-out edge has a free-fall time of $\sim$ 10$^7$ yr and will undergo little evolution over the next 5000 yr followed in the simulation. Our cut-out technique leads to the propagation of a rarefaction wave starting from the cut-out edge due to the vacuum boundary condition. However, the wave will only travel a distance of $c_s t$, where $c_s$ is the gas sound-speed ($\sim$ 2 km\,s$^{-1}$), and the time $t$ is 5000 yr. This corresponds to an insignificant distance of $\sim$ 10$^{-2}$ pc (2000 AU) from the cut-out edge, over two orders of magnitude smaller than the 10 pc box size.

 In addition, each remaining SPH particle is replaced with 256 child particles, each of which is placed randomly within the smoothing kernel of the parent particle.  The mass of the parent particle is then evenly divided amongst the child particles.  Each of these particles inherits the same chemical abundances, velocity, and entropy as the parent particle (see, e.g. \citealt{bromm&loeb2003}; \nocite{clarketal2011b} Clark et al. 2011b). This ensures conservation of mass, internal energy, and linear momentum.  After this process, each SPH particle in the new cut-out simulation has a mass of $m_{\rm sph}=7.2\times10^{-3}$ M$_{\odot}$, and the new resolution mass is $M_{\rm res} \simeq 0.4$ M$_{\odot}$, 
the approximate mass contained in one SPH kernel (\citealt{bateetal1995}).  
The cut-out and refinement is performed for the central star-forming region within each of the ten identified minihaloes.  
We then evolve the star-forming disks in these ten different minihaloes over 5000 yr of protostellar accretion.

\begin{figure}
\includegraphics[width=.45\textwidth]{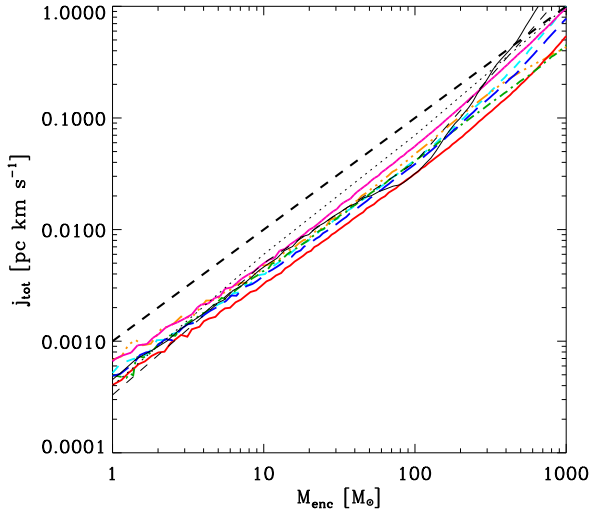}
 \caption{Angular momentum profile of gas within each minihalo, measured just prior to the formation of the first sink.  
The thick lines of various colors and linestyles represent 
six arbitrarily chosen
minihaloes from our suite.
The thin 
black 
lines denote angular momentum profiles found in separate cosmological simulations.  Solid 
black
line is taken from Stacy et al. (2010), dotted line from Yoshida et al. (2006), and dashed 
black
line from Abel et al. (2002).  Thick black dashed line shows an approximate powerlaw fit to these profiles, $j \propto M_{\rm enc}$.
Note the near convergence of each of these angular momentum profiles, including profiles obtained from varying cosmological realizations 
and differing techniques for hydrodynamics computation.
}
\label{amomprof}
\end{figure}

\subsection{Chemistry, Heating, and Cooling}

We utilize the same chemistry and thermal network as described in detail by \cite{greifetal2009} and used in \cite{stacyetal2012}.   Briefly, the code follows the abundance evolution of  
H, H$^{+}$, H$^{-}$, H$_{2}$, H$_{2}^{+}$, He, He$^{+}$, He$^{++}$, and e$^{-}$, as well as the three deuterium species D, D$^{+}$, and HD.  
All relevant cooling mechanisms, including H$_2$ collisions with  H and He as well as other H$_2$ molecules, are included.  The thermal network also includes cooling through  
H$_2$ collisions with protons and electrons, H and He collisional excitation and ionization, recombination, bremsstrahlung, and inverse Compton scattering off the cosmic microwave background.  

Further H$_2$ processes are also included to properly model gas evolution to high densities.  For instance, the chemistry and thermal network includes three-body H$_2$ formation and the concomitant H$_2$ formation heating, which become important at  $n \ga 10^8$ cm$^{-3}$.  
Three-body H$_2$ formation rates are uncertain, however, and \cite{turketal2011} determined that the variation in suggested rates leads to significant differences in the gas collapse at high density, including long-term disk stability and fragmentation.
For our simulations, we choose the intermediate rate from \cite{pallaetal1983}.
When $n \ga 10^9$ cm$^{-3}$, cooling through H$_2$ ro-vibrational lines becomes less effective as these lines grow optically thick.  We account for this using an escape probability formalism together with the Sobolev approximation (see \citealt{yoshidaetal2006,greifetal2011} for further details).

\subsection{Sink Particle Method}

SPH particles that reach densities of $n_{\rm max} = 10^{13}$ cm$^{-3}$ are converted into sink particles.  Sinks grow in mass by accreting surrounding gas particles that lie within a distance of  $r_{\rm acc}$, where 
 $r_{\rm acc} = L_{\rm res} \simeq 20$ AU, and

\begin{equation}
L_{\rm res}\simeq 0.5 \left(\frac{M_{\rm res}}{\rho_{\rm max}}\right)^{1/3} \mbox{\ .}
\end{equation}

Along with the criterion that the accreted particles must be within a distance $d  < r_{\rm acc}$ from the sink, we also require that the particles are not rotationally supported against infall onto the sink.  This corresponds to  
$j_{\rm SPH} < j_{\rm cent}$, 
where $j_{\rm SPH} = v_{\rm rot} d$ is the specific angular momentum of the gas particle, $j_{\rm cent} = \sqrt{G M_{\rm sink} r_{\rm acc}}$ is the angular momentum required for centrifugal support, and $v_{\rm rot}$ and $d$ are the
rotational velocity and distance of the particle relative to the sink.  
In addition, a sink accretes all particles which come within $d < r_{\rm min} = 4$\,AU.
These same criteria are used not only for gas particles but also for neighboring sink particles, so our algorithm allows for the merging of two sink particles. 
When the sink first forms, it immediately accretes the majority of particles within $r_{\rm acc}$,  giving it an initial mass of $M_{\rm res} \simeq 0.4$ M$_{\odot}$.  
Each time the sink grows, its position and velocity are set to the mass-weighted of that of the sink and the accreted particles.

After a particle becomes a sink, its density, temperature, and chemical abundances are no longer updated, though its position and velocity continue to evolve through gravitational interactions.  Each sink is instead held at a constant density  $n_{\rm max}$ and temperature of 1000 K.
The sink thus exerts a pressure on the surrounding particles, avoiding the formation of an artificial pressure vacuum around its accretion radius (see \citealt{brommetal2002,marteletal2006}).  

Once a sink is formed, nearly all gas particles near the sinks will be accreted before approaching the $d < r_{\rm min}$ criterion.  However, secondary sinks can come within this distance and become subject to strong N-body interactions.  In physical reality, however, the sub-sink region will contain not only the protostar but also surrounding gas and an accretion disk.  In addition, the protostars themselves can be very distended, with photospheres reaching $\sim$ 1 AU (e.g. \citealt{omukai&palla2003,greifetal2012}\nocite{smithetal2012}; Smith et al. 2012a).  
Our merging of all secondary sinks within  $d < r_{\rm min}$ therefore 
provides a rough treatment of unresolved sub-sink hydrodynamic effects which will mitigate those of pure N-body dynamics.

\begin{figure*}
\includegraphics[width=.8\textwidth]{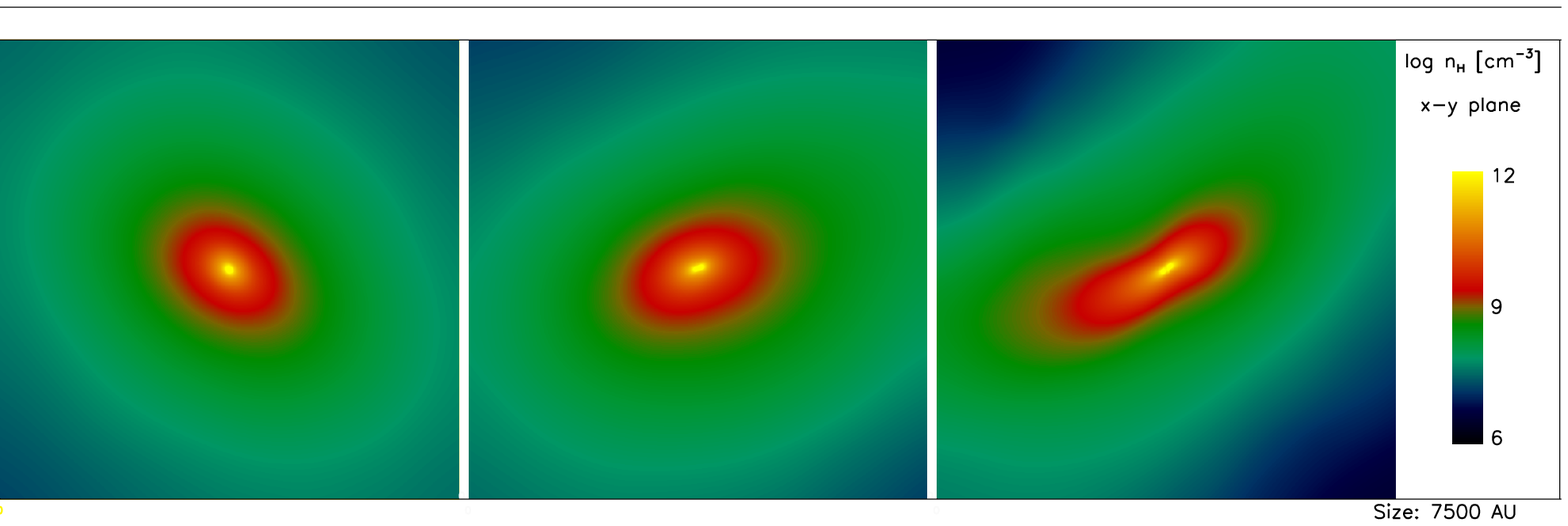}
\includegraphics[width=.8\textwidth]{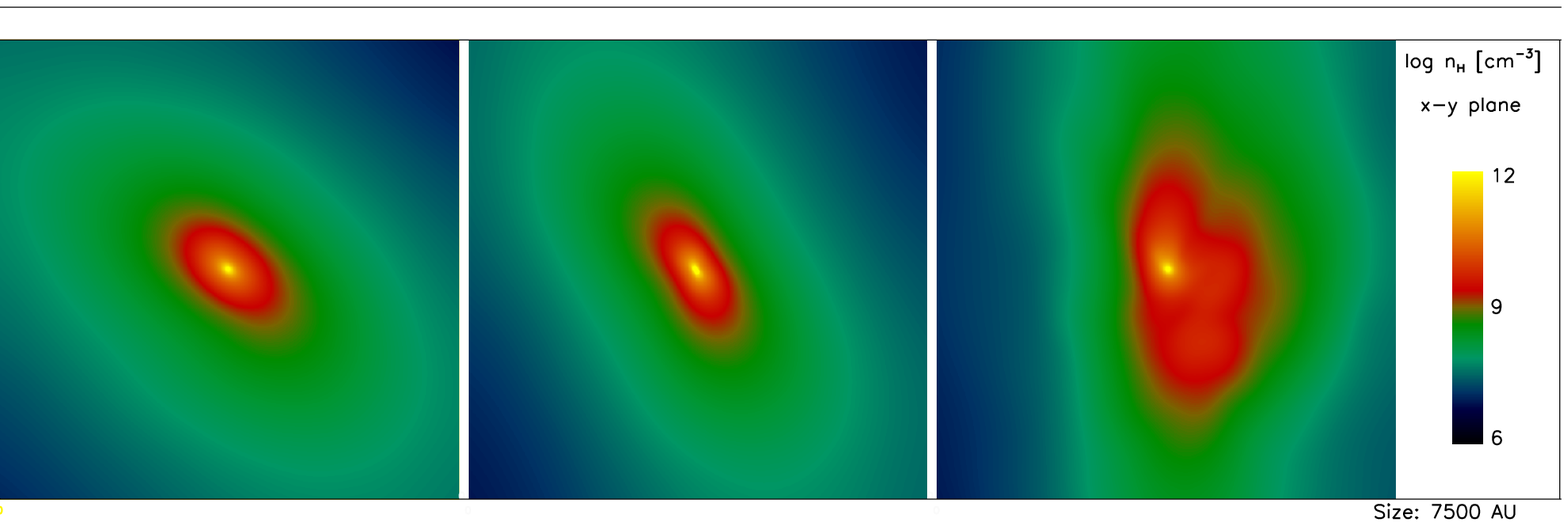}
 \caption{Density structure of the gas just prior to the formation of the first sink particle in six select minihaloes.
Each star-forming region has a similar structure which roughly follows that of an isothermal sphere.  However, the gas within differing minihaloes does vary somewhat in the degree of flattening.
}
\label{nh-morph}
\end{figure*}

\subsection{Identifying Binaries}

Once a protostellar multiple system has formed, we identify which pairs of sinks comprise a binary system.  For each minihalo we determine which pairs of sinks are gravitationally bound by comparing their specific potential and kinetic energies, $\epsilon_{\rm p}$ and $\epsilon_{\rm k}$.  A pair of sinks is considered a binary system if $\epsilon< 0$, where $\epsilon$ is the total orbital energy, 

\begin{equation}
\epsilon  = \epsilon_{\rm p} + \epsilon_{\rm k} \mbox{,}
\end{equation}

\begin{equation}
\epsilon_{\rm p} = \frac{-G(M_1 + M_2)}{r} \mbox{,} 
\end{equation}

\noindent and

\begin{equation}
\epsilon_{\rm k} = \frac{1}{2}v^2 \mbox{.} 
\end{equation}

\noindent $M_1$ and $M_2$ are the masses of the two sinks in question, $r$ is the distance between the sinks, and $v$ is the relative velocity between the sinks.   
Within each minihalo we find an average of between two and three binaries.

Each binary is then assigned an overall mass $M_{\rm bin} = M_1 + M_2$, a center-of-mass (COM) position, and a COM velocity. 
We next determine the semi-major axis, $a$, for each binary.  This is derived by equating the above definition of $\epsilon$ with the following energy equation:

\begin{equation}
\epsilon =  \frac{-GM_{\rm bin}}{2a} \mbox{.} 
\end{equation}

\noindent This yields

\begin{equation}
a = \frac{-G\left( M_1 + M_2 \right)}{2\left(\epsilon_{\rm k} + \epsilon_{\rm p}\right)} \mbox{.}
\end{equation}

We furthermore calculate the binary's mass ratio $q$, where

\begin{equation}
q = \frac{M_2}{M_1} \mbox{.}
\end{equation}

\noindent In this case $M_2$ corresponds specifically to the less massive of the binary members, and $M_1$ corresponds to the more massive, such that the value of $q$ is always between 
zero and one.

\begin{figure*}
\includegraphics[width=.8\textwidth]{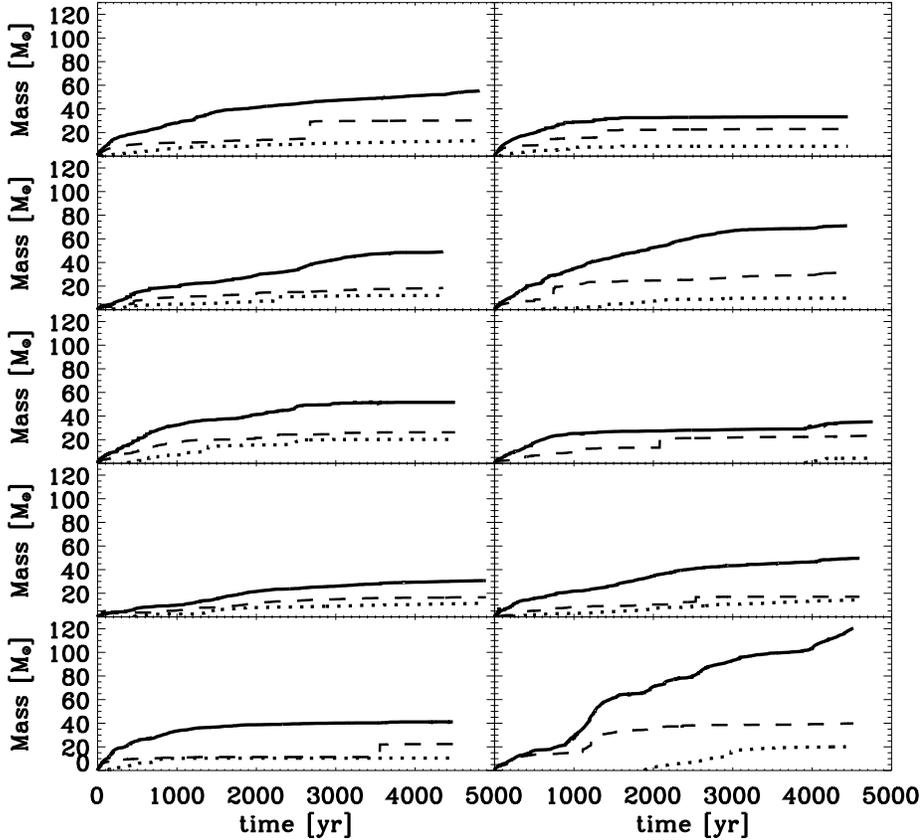}
 \caption{ History of the mass growth of the stellar systems in each minihalo.  Solid lines represent the total sink mass.  Dashed lines show the growth of the most massive sink.  Dotted lines depict the growth of the second-most massive sink.  Average accretion rates vary by over a factor of three between minihaloes.
}
\label{sinkmass}
\end{figure*}

\section{Results}

\subsection{Initial Minihalo Collapse}

The initial collapse of gas to protostellar density exhibits substantial similarity within each minihalo.  Fig. \ref{Tvsn} shows the thermal state of the star-forming gas within each minihalo just prior to the formation of the first maximally resolved sink, while Fig. \ref{mini_vel} shows their velocity profiles.  At high densities of $n >$ 10$^8$ cm$^{-3}$, H$_2$ three-body formation becomes rapid.  This leads to nearly isothermal evolution as increased H$_2$ cooling counteracts adiabatic heating while the gas condenses to $n > 10^{12}$ cm$^{-3}$.  
The minihaloes show a striking similarity in thermal evolution in this density range.
In the evolution of gas to the densities shown in Fig. \ref{Tvsn} ($n >$ 10$^6$ cm$^{-3}$), we also find that the HD cooling is generally unimportant compared to H$_2$ cooling,


Fig. \ref{amomprof} shows the angular momentum profile of the gas in the maximally resolved simulations just before the first sink has formed, measured with respect to the COM of all gas included in the 10 pc simulations.  These profiles all have roughly the same magnitude and powerlaw slope, such that $j(r) \propto M_{\rm enc}$, where $M_{\rm enc}$ is the enclosed gas mass within a given radius $r$ from the COM.
This convergence is found not only among minihaloes in our simulation, but also including those from differing cosmological realizations, performed with varying methods of computational hydrodynamics (\citealt{abeletal2002,yoshidaetal2006}).
These angular momentum distributions may additionally lead to high spin rates of individual protostars (e.g. \nocite{stacyetal2011}Stacy et al. 2011a; \citealt{stacyetal2013}).

\begin{figure*}
\includegraphics[width=.8\textwidth]{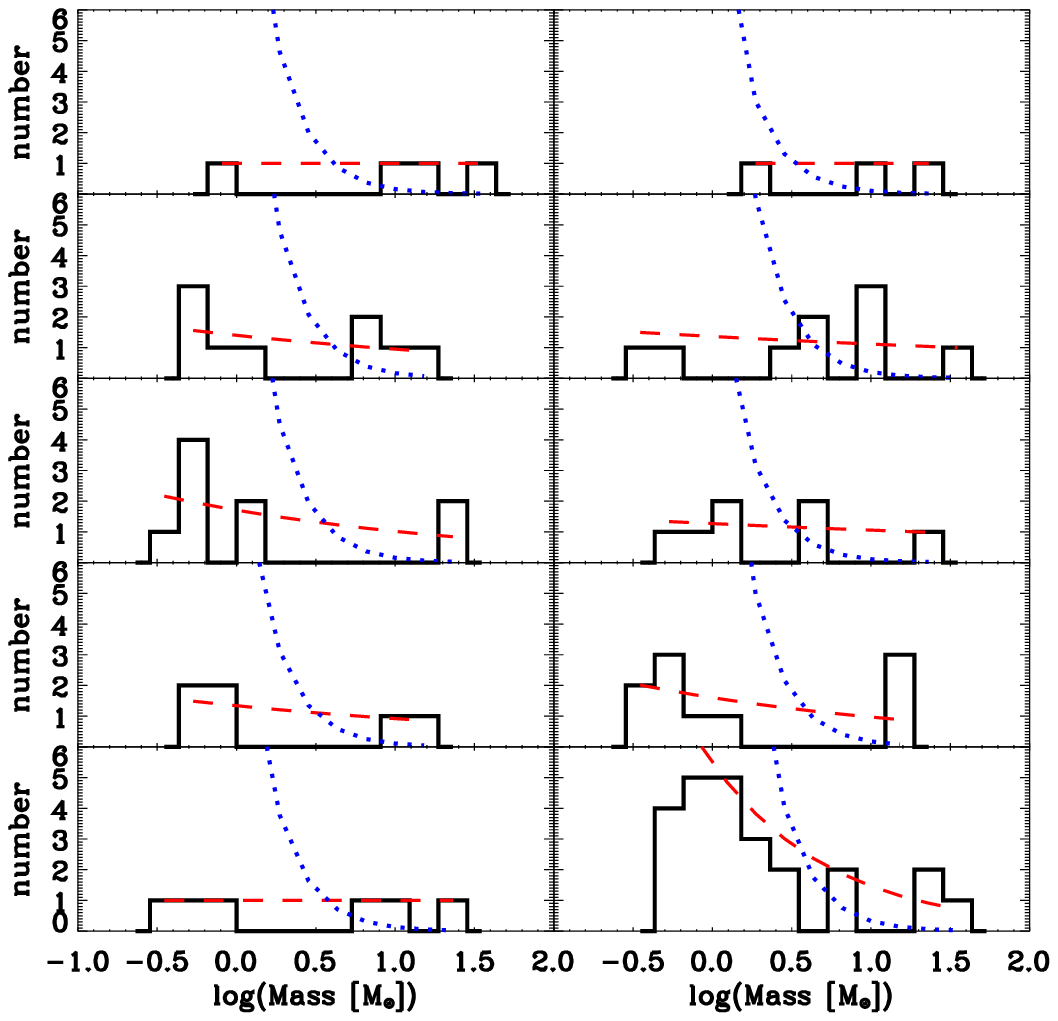}
 \caption{Distribution of sink masses in each minihalo after $\sim$ 5000 yr of accretion.  Dashed red lines depict powerlaw fits to each distribution,  ${\rm d}N/{\rm d}m \propto M_*^{-\alpha}$.  We find that $\alpha$ ranges from zero (Minihaloes 1, 2 and 9) to $0.57$ (Minihalo 10).  
 Dotted blue lines show an example distribution for $\alpha=2$, the minimum powerlaw that will  yield a top-heavy mass function.
For comparison, this is also similar to the observed Salpeter IMF of  $\alpha=2.35$.  
Each $\alpha=2$ distribution is normalized to have the same total mass as that found within the sinks of the corresponding minihalo.
}
\label{m_hist}
\end{figure*}

\begin{figure}
\includegraphics[width=.4\textwidth]{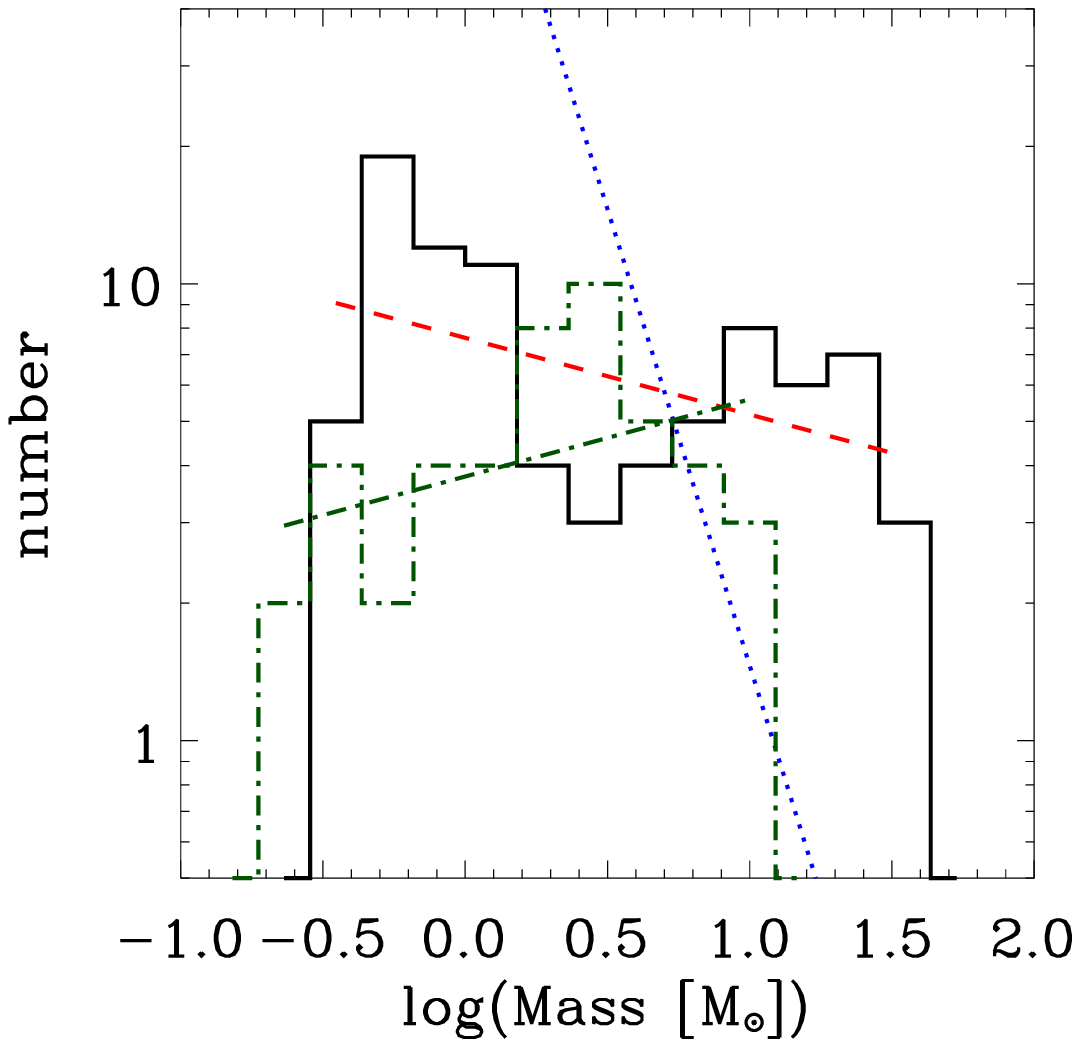}
 \caption{Total distribution of sink masses in all minihaloes.  Red dashed line depicts a powerlaw fit to the overall distribution,  ${\rm d}N/{\rm d}m \propto M_*^{-0.17}$.  
 Green dash-dotted lines depict the total mass distribution from the `merging sink' simulations of Greif et al. 2011, and a powerlaw fit to this distribution ($\alpha=-0.17)$.
For comparison, the blue dotted line again shows an example  $\alpha=2$ distribution, similar to the observed Salpeter IMF ($\alpha=2.35$).
}
\label{m_hist_tot}
\end{figure}

The density structure of the central gas can be seen in Fig. \ref{nh-morph} for a representative number of the minihaloes.  The similarity between minihaloes is again apparent, though some of the star-forming gas exhibits greater elongation and a more disk-like structure.
Similar variation is seen when we compare the velocity and turbulent structure in Fig. \ref{mini_vel}.  
We measured the rotational and radial velocities of each gas particle with respect to the COM of the densest gas, those particles with $n > 2\times10^{11}$ cm$^{-3}$.  The  rotational velocity $v_{\rm rot}$ and  radial velocity $v_{\rm rad}$ within each logarithmically-spaced
radial bin are then taken as the mass-weighted average of the individual particle velocities within each bin.    
In a similar fashion,
we measure $M_{\rm turb}$ over the same range of 
radial bins according to the following:
\begin{equation}
M_{\rm turb}^2 c_s^2 =\sum_{i} \frac{m_i}{M}\left(\vec{v}_i - \vec{v}_{\rm rot }  - \vec{v}_{\rm rad  }\right)^2 \mbox{,}
\end{equation}
where $c_s$ is the sound speed of the radial bin, $m_i$ is the mass of a gas particle contributing to the bin, and $M$ is the total mass of the bin.
Between 20 and 10,000 AU, both $v_{\rm rot}$ and $v_{\rm rad}$ remain at a few km s$^{-1}$, while the turbulence remains approximately sonic, never reaching more than twice the thermal velocity of the gas.  In each case, both radial and infall velocities are of the order of $\sim 1/2$ of the free-fall velocity $v_{\rm ff}$.  However, the relative magnitude of $v_{\rm rad}$ and $v_{\rm rot}$ varies, with $v_{\rm rad}$ dominating in some cases and  $v_{\rm rot}$ dominating in others.

After the central gas in each minihalo reaches the density threshold for sink formation, in all cases a stellar multiple system quickly forms.  This fragmentation is in agreement with the parameter study of \cite{machidaetal2008}.  They modeled their initial conditions as modified Bonnor-Ebert spheres, which well-describes collapsing and approximately isothermal gas like that within star-forming regions (e.g. \citealt{ebert1955,bonnor1956,stacyetal2010,greifetal2012}), and found the following fragmentation condition:

\begin{equation}
\Omega_0 > \Omega_{\rm crit} = 4\times10^{-17} \left(\frac{n_0}{10^3 \rm cm^{-3}}\right)^{2/3} \rm s^{-1} \mbox{,}
\end{equation}

\noindent where $\Omega_0$ and $n_0$ are the angular velocity and central number density of the initial core of the Bonnor-Ebert sphere. We determine $\Omega_0$ of the central core when $n_0$ is between $10^3$ cm$^{-3}$ and $10^4$ cm$^{-3}$, corresponding to  
$\Omega_{\rm crit}$ values between $\ga4\times10^{-17}$ s$^{-1}$ and  $\sim2\times10^{-16}$ s$^{-1}$.  We take a mass-weighted average of the angular velocity of all gas particles with  $n > 10^3$ cm$^{-3}$ and find in each case that $\Omega_0$ ranges between two and three orders of magnitude greater than $\Omega_{\rm crit}$.  The subsequent flattening and fragmentation of the dense gas are thus consistent with the criterion presented by \cite{machidaetal2008}.

\subsection{Mass Function}

The mass growth history of the sinks shows substantial variation among the ten minihaloes, as is shown in Figs \ref{sinkmass} and \ref{m_hist}.  At the end of the simulations, the total mass accreted by all sinks ranges from $30-120$\,M$_{\odot}$ for a given minihalo 
 (Fig. \ref{sinkmass}).    
The total mass of gas within each minihalo is typically $\la10^5$ M$_{\odot}$, implying a star-formation efficienty of $\sim10^{-3}$. 
The distribution of sink masses after 5000 yr of accretion can be seen for the individual minihaloes in Fig. \ref{m_hist}, where the variation between minihaloes is even more striking.  In particular, while some minihaloes have only one or two sinks with $M_* < 1$ M$_{\odot}$, this number can be as large as nine.  We find a range of powerlaw fits to the relation ${\rm d}N/{\rm d}m \propto M_*^{-\alpha}$, where $N$ is the number of sinks that lie within a logarithmically scaled mass bin.
This range extends from $\alpha=0$ for Minihaloes 1, 2 and 9, to $\alpha=0.57$ for Minihalo 10.
Each minihalo has only a handful of stars, so the powerlaw fits presented in Fig. \ref{m_hist} are meant only to provide rough 
comparisons between minihaloes.
When we fit to a larger sample of sinks by
considering the distribution across all minihaloes (Fig. \ref{m_hist_tot}), we find ${\rm d}N/{\rm d}m \propto M_*^{-0.17}$.

Our choice of fitting a powerlaw function to the sink mass distirbutions also allows us to directly compare our results with other initial mass functions (IMFs).
Unlike the observed present-day IMF, which is characterized by a Salpeter slope of $\alpha = 2.35$ towards the high mass end, our simulations yield a top-heavy mass function in that the majority of the protostellar mass resides within the most massive protostars.  
Top-heaviness applies to mass functions with $\alpha < 2$, since this is the requirement for the average stellar mass to be dominated by the upper limit of the mass range (e.g. \citealt{bromm2012}).
If we consider smaller values of $\alpha$ to be more top-heavy, then our fitted mass function is less top-heavy than that found after $\sim 1000$ yr of sink accretion in \cite{greifetal2011}, who find $\alpha \la 0$.   

Including radiative feedback would likely modify our values for $\alpha$, though the magnitude of the modification is uncertain.  While feedback does not prevent fragmentation (e.g. \citealt{smithetal2011, stacyetal2012}), 
heating of the gas may suppress low-mass fragmentation to yield smaller $\alpha$ values closer to zero (i.e., a flatter spectrum). 
\cite{greifetal2011} resolved smaller mass scales and again found smaller values of $\alpha$. However, following such a resolved simulation for longer than their 1000 yr may reveal a population of very low-mass stars that arises at later times, instead increasing $\alpha$.
Our simulations do not resolve down to the opacity limit of fragmentation, the scale on which a small $\sim10^{-2}$ M$_{\odot}$ hydrostatic core first forms (e.g. \citealt{rees1976,omukai&nishi1998,greifetal2012}). We may therefore underestimate the number of such low-mass protostellar fragments that form.

 \begin{figure*}
\includegraphics[width=.4\textwidth]{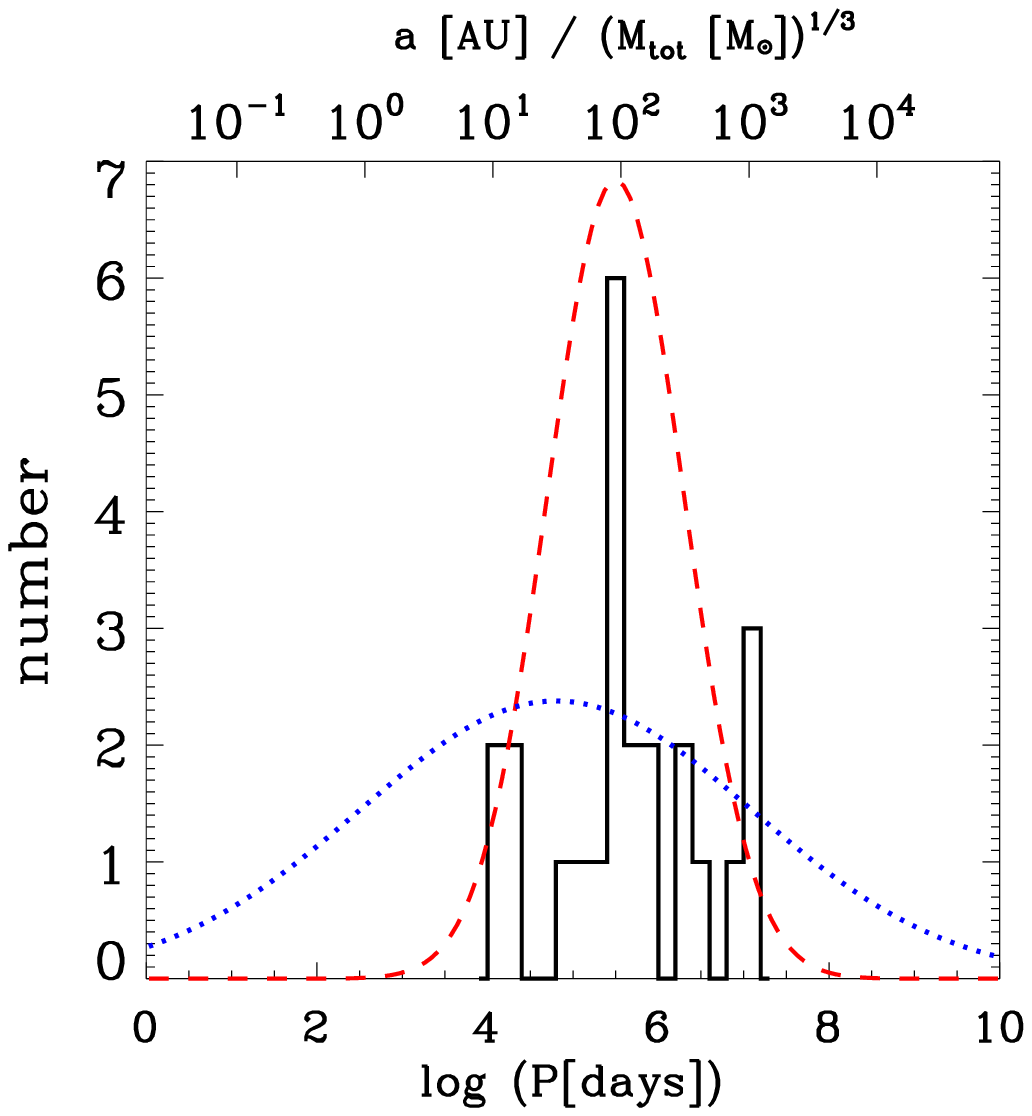}
\includegraphics[width=.4\textwidth]{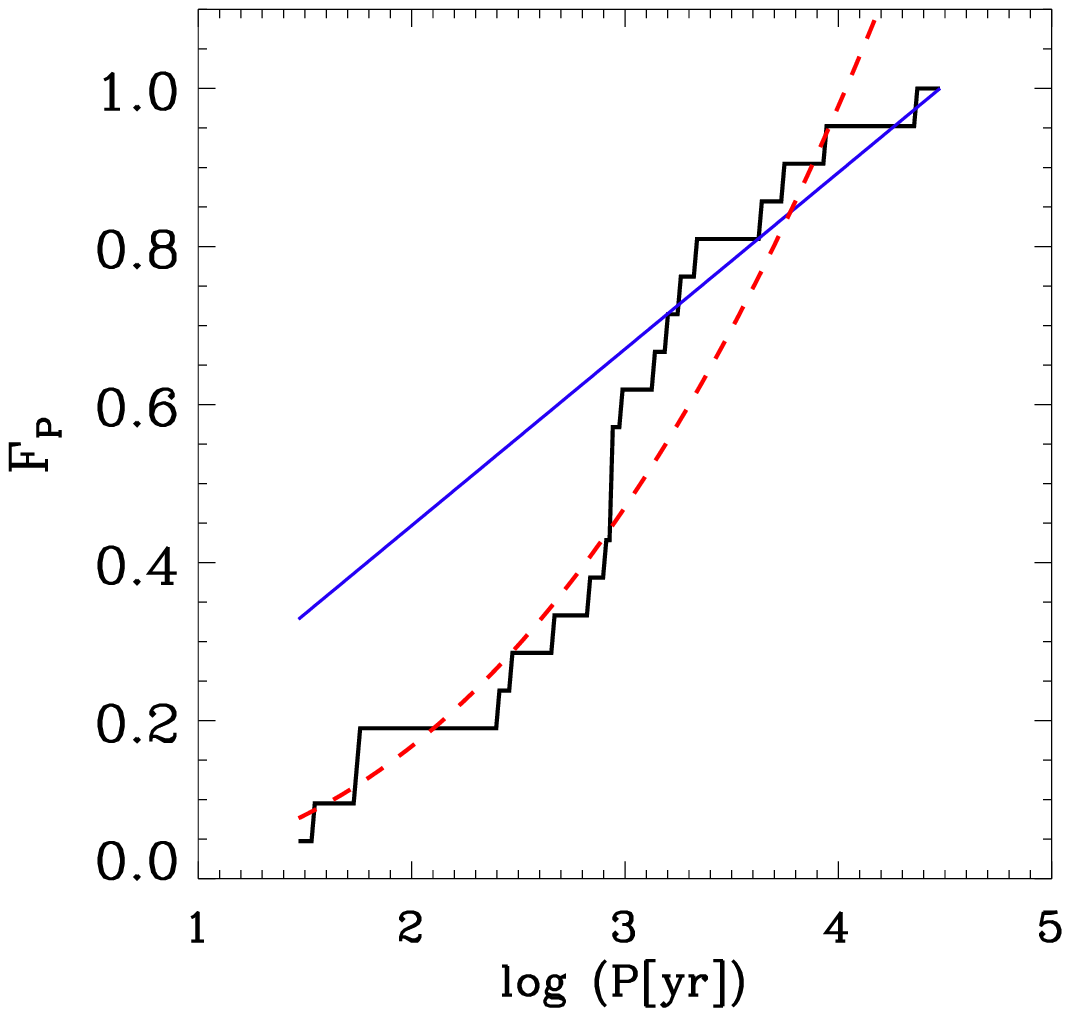}
 \caption{
 {\it Left:} Distribution of the log of the period, $P$, for the binaries found within each minihalo.  Note that units of $P$ is in days.  
 Red dashed line shows a Gaussian fit to the distribution.
Blue dotted line shows an example fit to data for solar-type stars (Duquennoy \& Mayor 1991).
 The distribution of the binary periods is peaked at 10$^{5.5}$ days, or 870 yr.
{\it Right:} Cumulative distribution $F_{\rm P}$ of the orbital period of the binaries identified in all minihaloes.  Red dashed line shows a powerlaw fit, 
$F_{\rm P} \propto {\rm log}(P)^{2.54}$, while the blue line shows a sample {\"O}pik distribution for comparison.
}
\label{Fp}
\end{figure*}

\subsection{Binary Fraction and Merger Rate}

Every minihalo contains at least one binary, and on average each halo hosts between two and three binaries.
We define the binary fraction as

\begin{equation}
f_{\rm B} = \frac{B}{S + B} \mbox{,}
\end{equation}

\noindent where $S$ and $B$ are the number of single and binary systems, respectively.  Summing over all sinks remaining in each minihalo at the end of the simulations, we have $S=42$ and $B=24$, yielding 
$f_{\rm B} = 0.36$. Calculated slightly differently, this means any individual star has a probability $2B/(S + 2B)$=53\% of having a binary companion.
This is a substantial fraction, and similar to the range of massive star binary fractions observed in various clusters in the Milky Way (e.g. \citealt{sana&evans2011,kiminki&kobulnicky2012}).

While a total of 217 sinks form in our suite of minihaloes, 90 remain after 5000 yr.  More than half (59\%) of the sinks are therefore lost to mergers.  This is similar to the merger rate found in \cite{greifetal2012}.  However, it will require higher-resolution simulations to determine which of these mergers were actually unresolved but long-lived tight ($a < 20$ AU) binaries.  The results of \cite{greifetal2012}, as well as our criterion that only non-rotationally supported sinks are merged, support the likelihood that a large number of these sink mergers were indeed true protostellar mergers.  

\begin{figure*}
\includegraphics[width=.4\textwidth]{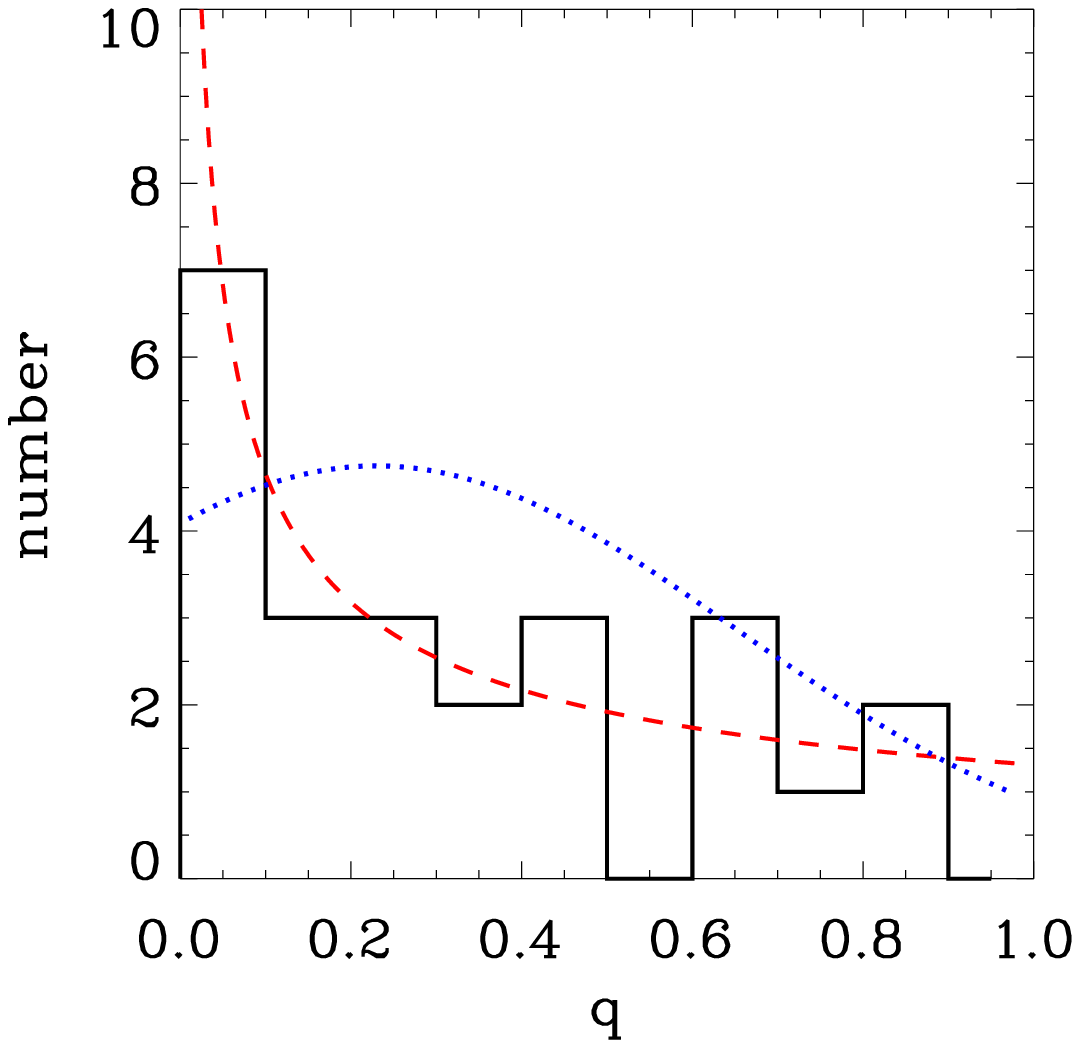}
\includegraphics[width=.4\textwidth]{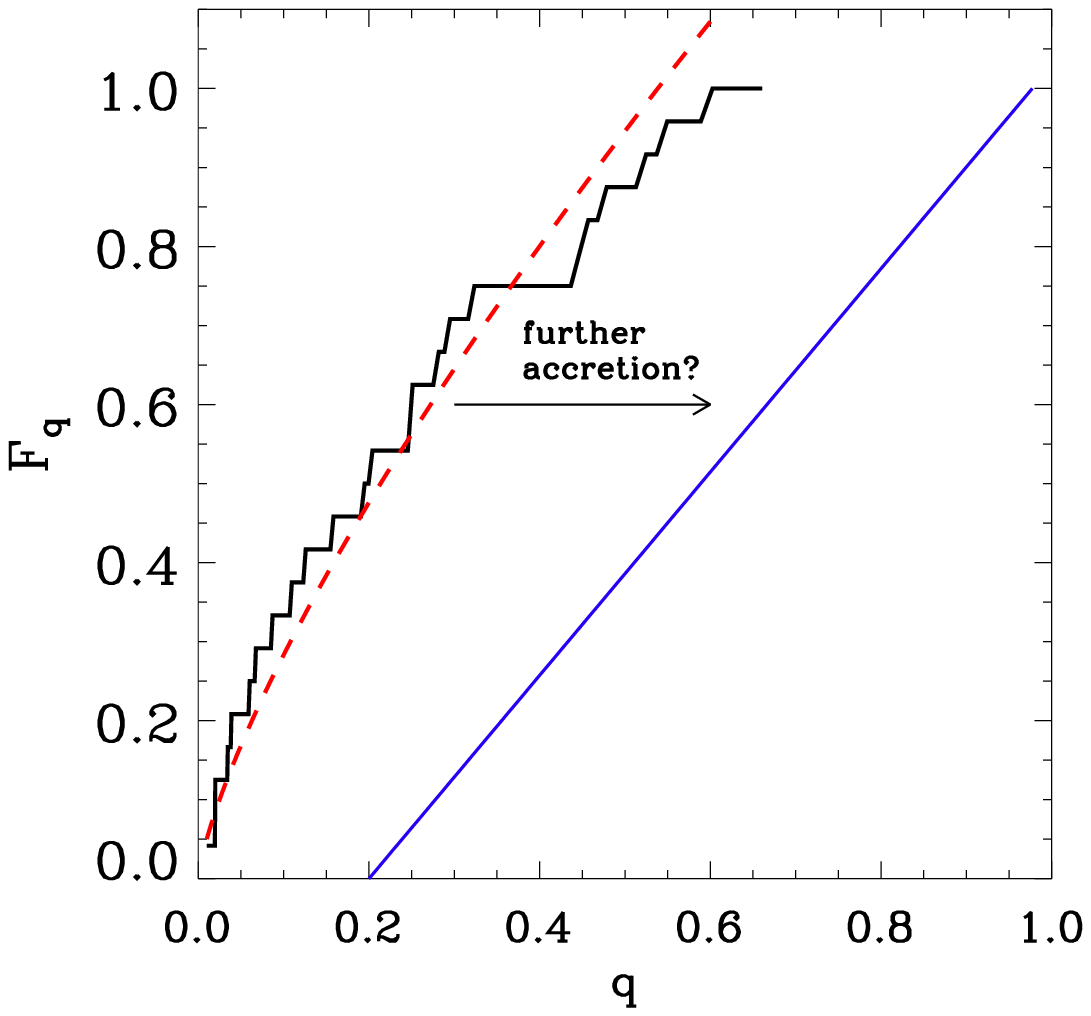}
 \caption{
{\it Left:} Distribution of the mass ratio $q$ between the members of binaries located in each minihalo.  Red line is the powerlaw fit $dN/dq \propto q^{-0.55}$.
Blue dotted line shows an example fit to data for solar-type stars (Duquennoy \& Mayor 1991).
The average value of $q$ is 0.34.
{\it Right:} Cumulative distribution $F_{\rm q}$ of the mass ratio of all binaries in our simulation.  Red dashed line is a powerlaw fit, $F_{\rm q} \propto q^{0.75}$.  The blue line depicts a uniform distribution down to $q=0.2$, as observed in massive star clusters in the Milky Way  (e.g. Sana \& Evans 2011). 
}
\label{q_hist}
\end{figure*}

\subsection{Orbital Period and Mass Ratio Distribution}

The characteristics of the binary distribution are shown in Figs 
 \ref{Fp}  and \ref{q_hist}. 
So that we may directly compare with Gaussian fits made to observed binary period distributions, we fit our period distribution to a Gaussian, as well. 
 The distribution of the binary periods is peaked at 10$^{5.5}$ days, or 870 yr.  In comparison with nearby solar-type stars (blue dotted line in Fig. \ref{Fp}), this peak is shifted to longer periods.
For instance, (\citealt{duquennoy&mayor1991}) observed a $P$ distribution that is peaked at 10$^{4.8}$\,days, or 170 yr.
This corresponds to a distribution of the semi-major axis, $a$, that is peaked around $\sim$ 250 AU, with an average value of 760 AU.  
In Fig. \ref{Fp} we also show the related cumulative distribution of the binary periods, $F_{\rm P}$, normalized such that $F_{\rm P}(P<\infty) = 1$.  
We display a best-fit powerlaw for this distribution, $F_{\rm P} \propto {\rm log}(P)^{2.54}$, and
compare to a `{\"O}pik distribution' (\citealt{opik1925}), where $F_{\rm P} \propto {\rm log}(P)$.  {\"O}pik's law is consistent with observations  of massive star binaries
(e.g. \citealt{kouwenhovenetal2007,sana&evans2011,kiminki&kobulnicky2012}).  However, the fit found from our simulation is significantly steeper.  As will further be discussed in Section 4, there are multiple reasons for this.  
The spatial resolution of the simulations was limited, so the distributions cannot  sample orbits with $a < 20$ AU.  This contributes to the apparent lack of hard binaries.  In addition, the binaries have only evolved for a few thousand years, and
their orbits may evolve and harden over longer periods of time.

The distribution of mass ratio $q$ (Fig. \ref{q_hist}) goes as $dN/dq \propto q^{-0.55}$, while for  the cumulative distribution we find  a best-fit powerlaw of $F_{\rm q} \propto q^{0.75}$.  
The average value of $q$ is 0.35.
We compare our $q$ distribution to that found for solar-type stars by \cite{duquennoy&mayor1991}, shown as the blue dotted line in Fig.~\ref{q_hist}.  They observed a $q$-distribution peaked at 0.23, while our distribution is peaked at $q<0.1$, so our simulations have a comparatively high fraction of low-$q$ binaries.
 
Our rate of low-$q$ binaries is also high when compared with more massive star clusters in the Milky Way.  For instance,  \cite{kiminki&kobulnicky2012} find that 
$dN/dq \propto q^{0.1}$, while our fit falls off much more steeply and is slightly outside their range of error.  
\cite{sana&evans2011} present observations which yield a uniform distribution of $q$ down to $q=0.2$. This corresponds to the cumulative distribution shown as the blue line in the right panel of Fig.~\ref{q_hist}.

If the sink particles of the simulations continued to accrete mass for longer periods of time, it is possible that the $q$ distribution would shift towards a larger percentage of binaries with nearly equal mass.
As described in e.g. \cite{bate&bonnell1997}, if gas with sufficient angular momentum flows onto a binary star, it will accrete preferentially onto the less massive companion.  Angular momentum conservation prevents rotating gas from reaching the larger star without the aid of, e.g., gravitational torques.  The gas instead remains at some radius beyond the more massive star where it is more easily captured by the less massive companion, resulting in a tendency for $q$ to evolve towards a value of one.  This is similar to the fragmentation-induced starvation described by \cite{petersetal2010b}.  In their simulations of star-forming molecular clouds, they found that gas inflowing through a stellar disk was accreted by lower-mass stars before it could reach the more massive and centrally-located star.

\begin{figure*}
\includegraphics[width=.45\textwidth]{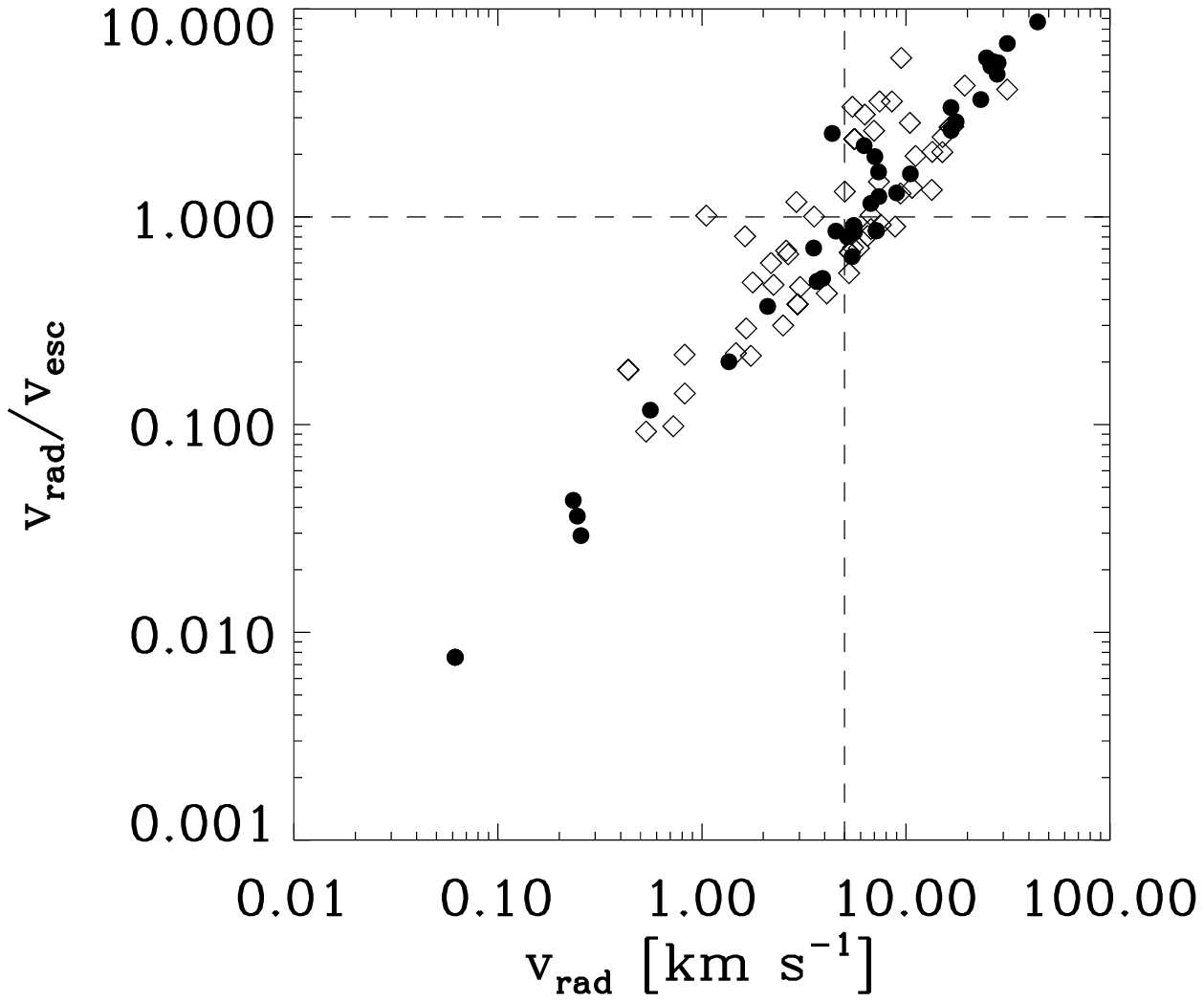}
\includegraphics[width=.45\textwidth]{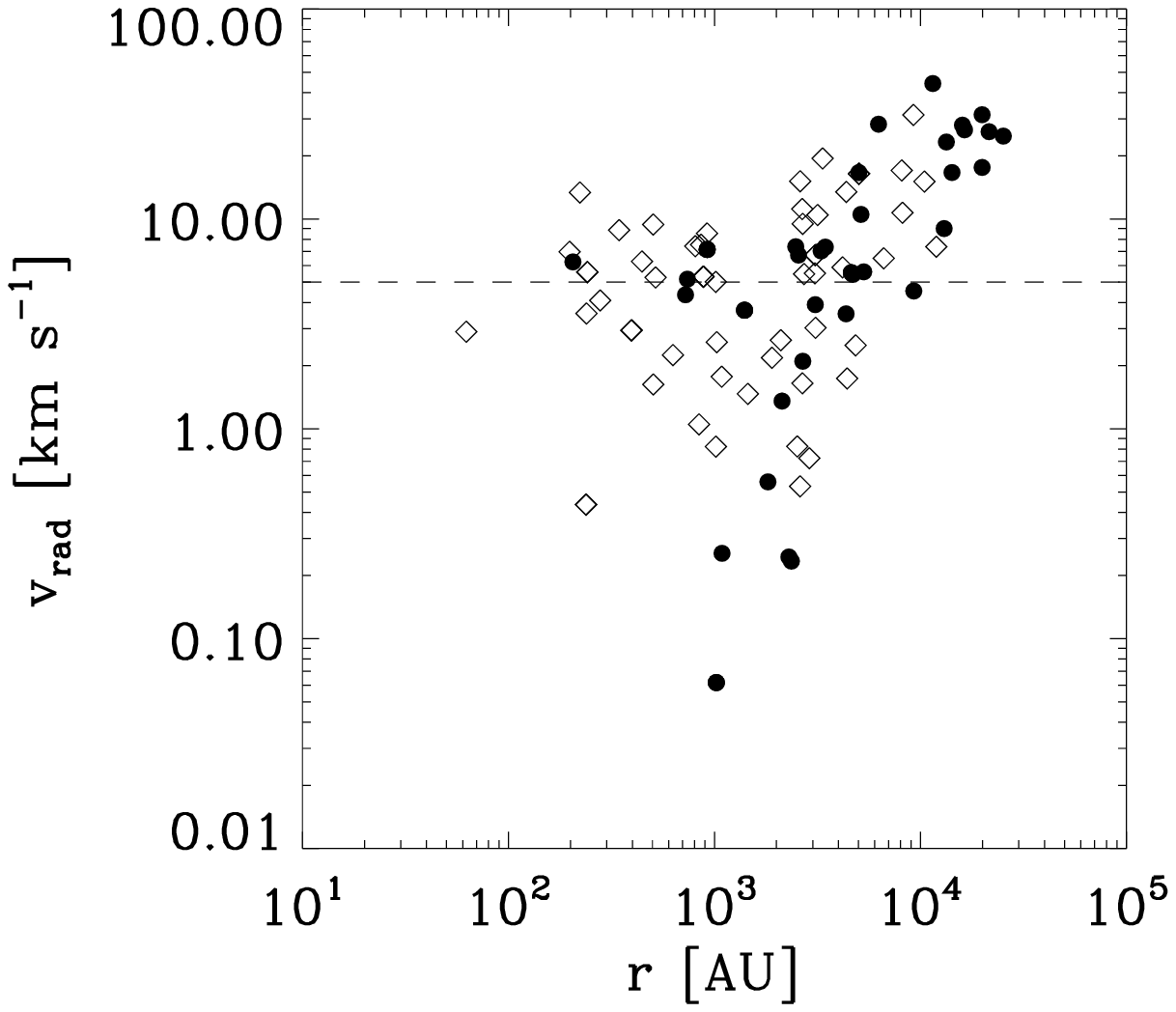}
 \caption{
 {\it Left:} Ratio of $|v_{\rm rad}|$ to $v_{\rm esc}$ for each sink particle.  Sinks with $M_* < 1$ M$_{\odot}$ are denoted with filled circles, while diamonds represent sinks with $M_* > 1$ M$_{\odot}$.
 Horizontal dashed line shows where $|v_{\rm rad}/v_{\rm esc}| =1$.  Sinks above the horizontal dashed line are able to escape from the stellar disk.  The vertical dashed line denotes $v_{\rm esc, halo}$, the velocity necessary to escape the minihalo.  
 {\it Right:} Variation of $|v_{\rm rad}|$ with radius.  Dashed line again shows $v_{\rm esc, halo}$.
}
\label{v_esc}
\end{figure*}

\subsection{Protostellar Ejection Rate}

Of the total of 90 sinks remaining at the end of our simulation, 39 have radial velocities that are greater than their escape velocities (Fig. \ref{v_esc}), where we define $v_{\rm esc}$ as

\begin{equation}
v_{\rm esc} = \sqrt{\frac{G M_{\rm enc}}{r}} \mbox{,}
\end{equation}

\noindent where $M_{\rm enc}$ is the mass enclosed between the COM and the sink particle at distance $r$ from the 
COM.  The COM is determined using all gas and sink particles with $n > 10^{10}$ cm$^{-3}$.
Roughly  half (43\%) of the sinks can thus leave the stellar disk, similar to the rate found in, e.g. \cite{greifetal2011} in their simulations of primordial star formation and \cite{bateetal2003} in their numerical studies of present-day, low-mass star formation.  

In addition, we define a second escape velocity, that required to exit the minihalo:

\begin{equation}
v_{\rm esc, halo} = \sqrt{\frac{G M_{\rm halo}}{r_{\rm halo}}} \sim 5 \rm \,km\,s^{-1} \mbox{,}
\end{equation}

\noindent where we define $M_{\rm halo} = 4 \times 10^5$ M$_{\odot}$ and $r_{\rm halo} = 80$ pc, the average values for the minihaloes of our suite.
Whether $v_{\rm esc}$ or $v_{\rm esc, halo}$ is larger varies depending on the sink and its location.  A total of 53 sinks have $v_{\rm rad} > v_{\rm esc, halo}$, which is greater than the 39 which can escape from the stellar disk.  Thus, the gravity of the stellar disk provides the stronger escape condition.  In the majority of cases, for a sink to have sufficient velocity to escape the disk, it must already have sufficient velocity to escape the minihalo (left panel of Fig. \ref{v_esc}).  For a sink traveling at $v_{\rm esc, halo}$, the time to cross the virial radius of the minihalo is $\sim 2\times10^7$ yr.  Note that within this time the minihalo is not expected to undergo significant mass growth itself, no more than doubling in mass (e.g. Stacy et al. 2011b\nocite{stacyetal2011b}, \citealt{wechsleretal2002}).  Thus, $v_{\rm esc, halo}$ will not increase by more than 1-2 km s$^{-1}$.

Of the ejected protostars, 18 are of mass $M_* < 1$~M$_{\odot}$ and thus have a possibility of surviving to the present day (Clark et al 2011b).  
This depends, however, on whether the stars will accrete more mass at a later time and thus become shorter lived.  
Identifying them as Pop~III stars would require that, as they become incorporated into the Milky Way or nearby dwarf galaxies, they will not accrete any enriched material that will remain in their atmospheres and mask them as Pop II stars 
(e.g. \citealt{frebeletal2009,johnson&khochfar2011}).   



\begin{figure*}
\includegraphics[width=.85\textwidth]{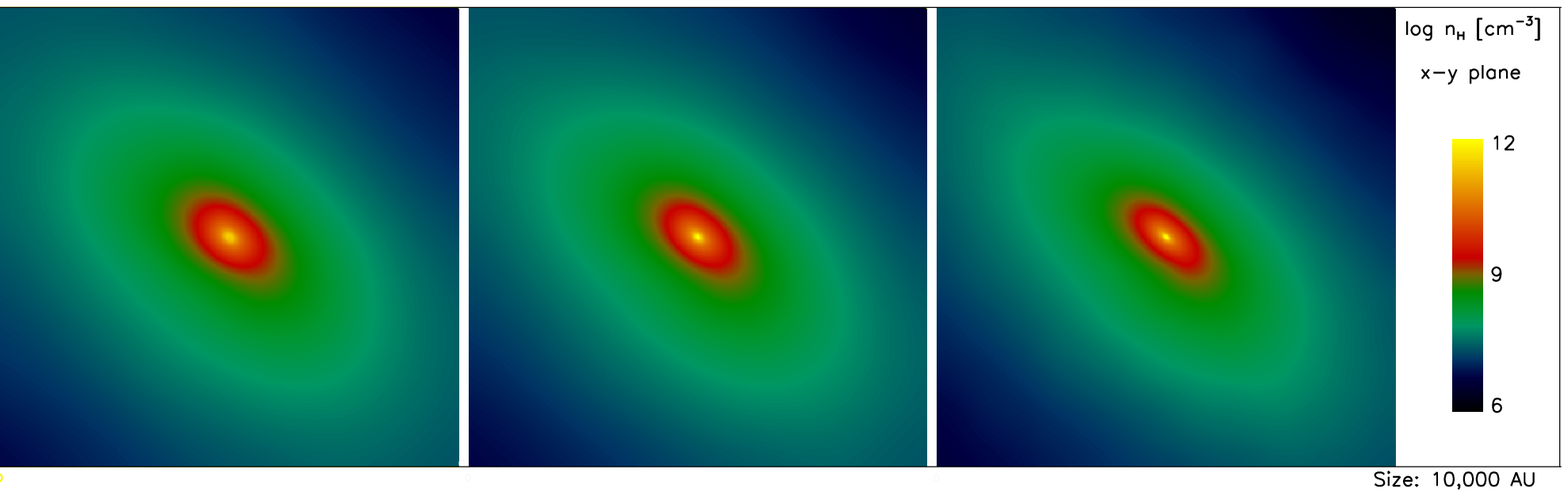}
\includegraphics[width=.85\textwidth]{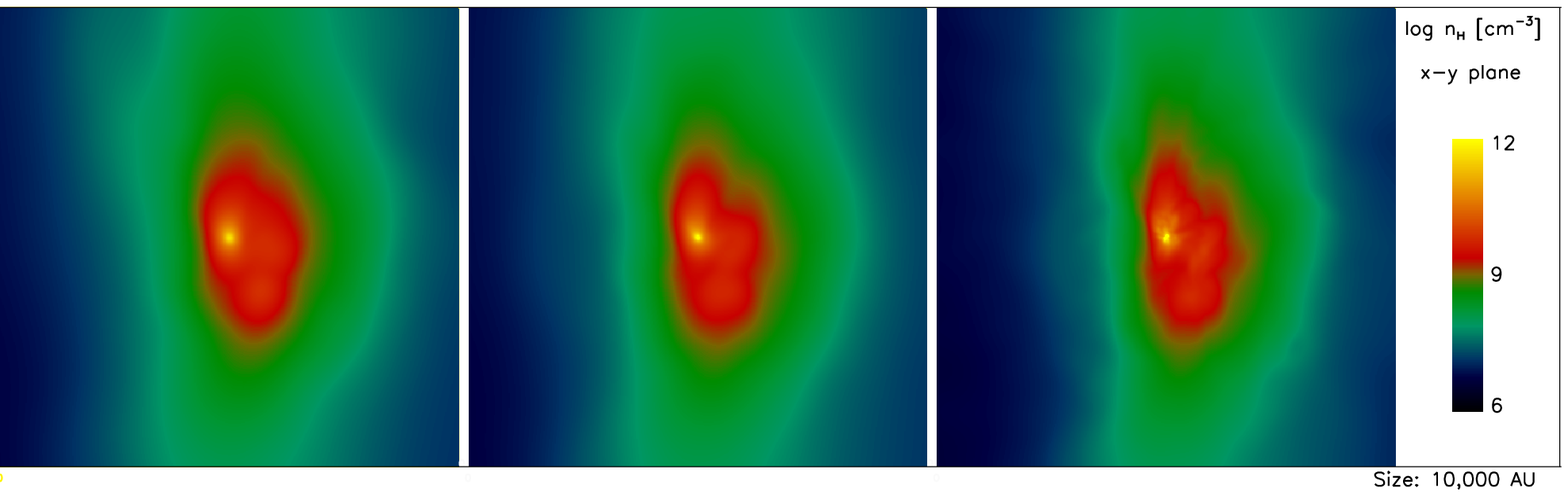}
 \caption{
 {\it Top:} Morphology of the LR case of Minihalo 7 (left), the original case (middle), and the HR case with extra refinement (right).
 {\it Bottom:} Same as top panels, but for Minihalo 10.
 The overall morphology is again similar over this range of scales, though more small-scale clumpiness is resolved in the  Minihalo 10 HR case.
}
\label{morph_comp}
\end{figure*}


\section{Resolution Study}
Here we describe how our results vary with resolution. In our original maximally resolved simulations (Section 2.2), we required that $M_{\rm res} \le M_{\rm Jeans} \sim 700 (T/200 {\rm K})^{3/2} (n/10^4 \rm cm^{-3})^{-1/2}$  M$_{\odot}$.  For our maximum density of $10^{13}$ cm$^{-3}$, which has a temperature of $\sim$ 1500 K, gas incorporated into sink particles has $M_{\rm Jeans} = 0.4$ M$_{\odot}$.  However, if we instead use the resolution criterion presented in \cite{turketal2010}, $M_{\rm res}$ will be reduced.  They employ a temperature of $T_{\rm min} = 200$ K even if the gas has not cooled to temperatures this low.  In our case, the \cite{turketal2010} criterion would yield  $M_{\rm res} \le M_{\rm Jeans} = 0.02$  M$_{\odot}$.

To satisfy this more stringent resolution criterion, we resimulate Minihaloes 7 and 10, but with an extra refinement step (the high-resolution 'HR' case).  Minihaloes 7 and 10 contain the most slowly and most rapidly accreting star-forming clusters, respectively.   When the gas within these minihaloes has reached 10$^8$ cm$^{-3}$, we further split each particle into 32 child particles in the same way as described in Section 2.2.  This yields $m_{\rm sph} = 2 \times 10^{-4}$ M$_{\odot}$ and $M_{\rm res} = 0.01$ M$_{\odot}$.  
We furthermore compare with a less-resolved study of Minihaloes 7 and 10, in which we use half as many particles as in the original cases.  To initialize the low-resolution (LR) case, in the final refinement step described in Section 2.2, we replace the SPH particles with 128 instead of 256 child particles.  
The mass of the gas particles in the LR minihaloes is $m_{\rm SPH} =0.014$ M$_{\odot}$, while $M_{\rm res} = 0.8$ M$_{\odot}$.
Fig. \ref{morph_comp} shows that the overall morphology is very similar for each of these resolution levels.  However, in Minihalo 10 some small-scale clumpiness is resolved in the HR case that is not visible in the other two cases.

\begin{figure}
\includegraphics[width=.45\textwidth]{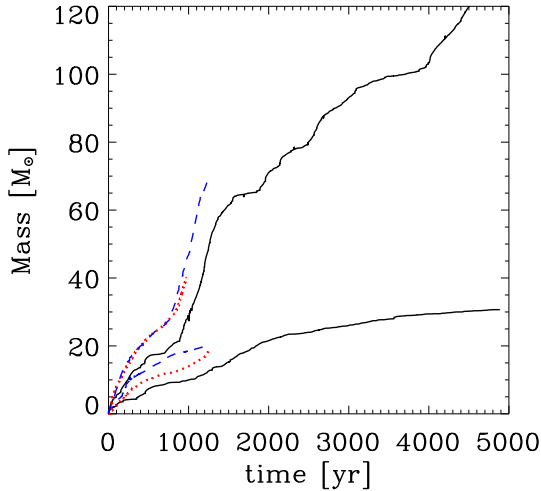}
 \caption{  
Growth of total sink mass over time.  Lower three lines depict the total mass of sinks within Minihalo 7, while upper three lines show that of Minihalo 10.  Black lines represent the sink growth within the original minihaloes, dotted red lines represent the sink growth when including extra refinement (HR case), and blue dashed lines depict the growth in the LR case.
}
\label{sinkmassE}
\end{figure}

\begin{figure*}
\includegraphics[width=.45\textwidth]{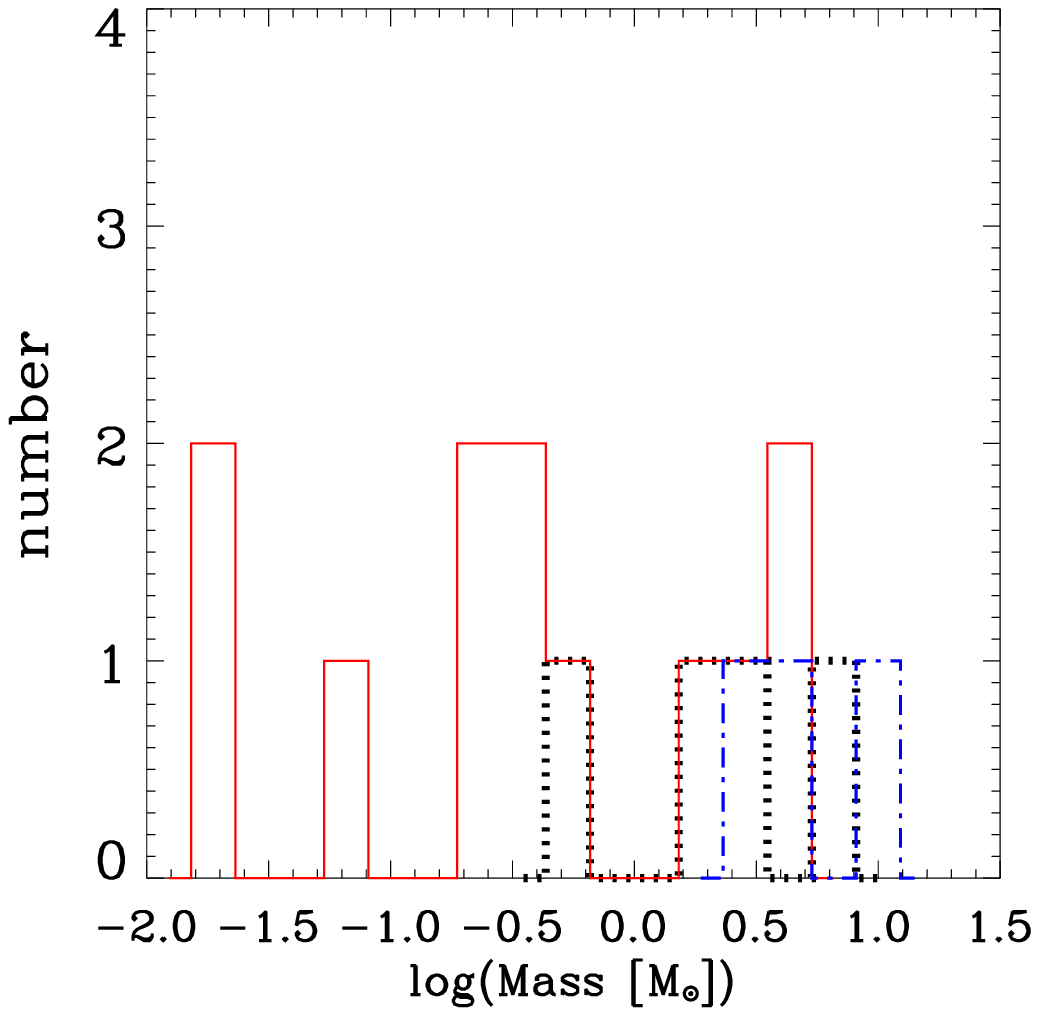}
\includegraphics[width=.45\textwidth]{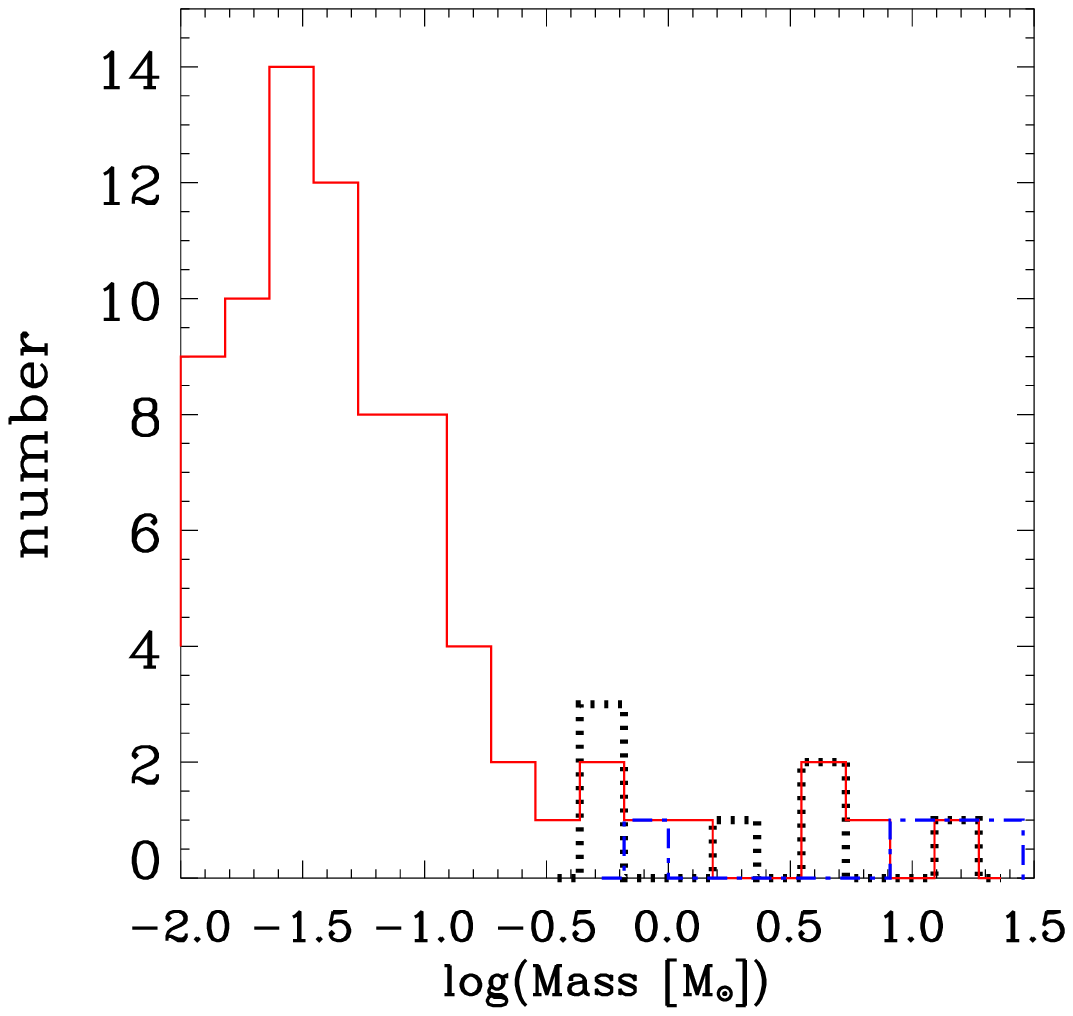}
 \caption{
{\it Left:}  Distribution of sink masses at 1000 yr in Minihalo 7.  Black dotted line represents the original minihalo, 
while red solid line represents the HR case and blue dash-dotted lines are from the low-resolution case.
{\it Right:} Distribution of sink masses at 1000 yr in Minihalo 10.  Lines have the same meaning as in the left panel.
}
\label{m_histE}
\end{figure*}

We also test how our results depend upon the scale at which sink particles are inserted.  In the more highly refined versions of Minihaloes 7 and 10, we insert a sink particle at the same density threshold $n_{\rm max} = 10^{13}$ cm$^{-3}$. The smaller $M_{\rm res}$ of these simulations leads to a corresponding sink accretion radius of $r_{\rm acc} = 6.5$ AU.  
In addition, we follow the sink accretion history of the LR cases, where  we insert sinks at densities of $n_{\rm max} = 10^{12}$ cm$^{-3}$ and employ an accretion radius of $r_{\rm acc} = 50$ AU. 
We follow the high and LR sink accretion for 1000 yr.  We compare the resulting mass growth in Fig. \ref{sinkmassE} and the distribution of sink masses in Fig. \ref{m_histE} at $t_{\rm acc} = 1000$ yr.

For Minihalo 7, we see that increasing resolution leads to a similar but enhanced sink growth history, where the total sink mass is $\sim$ 13 M$_{\odot}$ for the high-resolution case and 9 M$_{\odot}$ in the original minihalo (red line versus black line in Fig. \ref{sinkmassE}).  As is apparent in Fig. \ref{m_histE}, the mass distribution above $\sim$ 0.5 M$_{\odot}$ is similar, though the high-resolution case now includes a greater number of sinks with mass $\la$ 0.5 M$_{\odot}$, as well as three times the total number of sinks overall.  This is as expected since the original minihalo had a larger resolution mass of  $M_{\rm res} = 0.4$  M$_{\odot}$.
In contrast,  the low-resolution case  has a similar number of sinks as in the original case (see Table \ref{tab1}), but again with a greater total mass of $\sim$ 17 M$_{\odot}$ (blue line in Fig. \ref{sinkmassE}).  

Similarly, in Minihalo 10, the low-resolution case has greater total sink mass as in the original case for most of the accretion history (e.g. 30 versus 50 M$_{\odot}$ at 1000 yr).  The high-resolution case is in between the other two, with $\sim$ 40 M$_{\odot}$.  In addition, Fig. \ref{m_histE} shows that, as expected, each level of finer resolution allows the resulting IMF to extend to lower masses.  This is especially apparent for Minihalo 10, where a significant population of low-mass ($\la$ 1 M$_{\odot}$) sinks appears in the high-resolution case, though the distribution at higher masses is still similar for all three cases.

\begin{figure*}
\includegraphics[width=.45\textwidth]{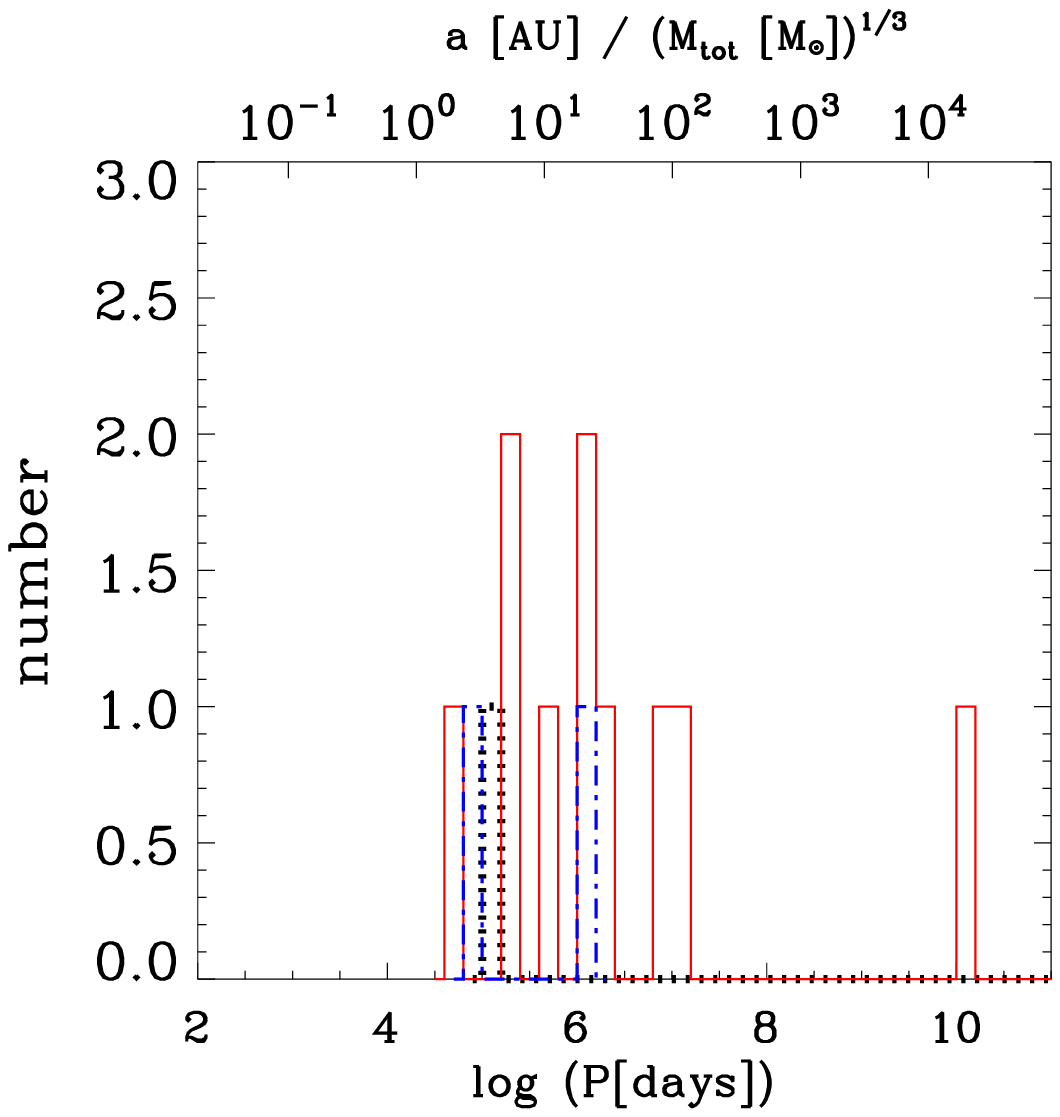}
\includegraphics[width=.45\textwidth]{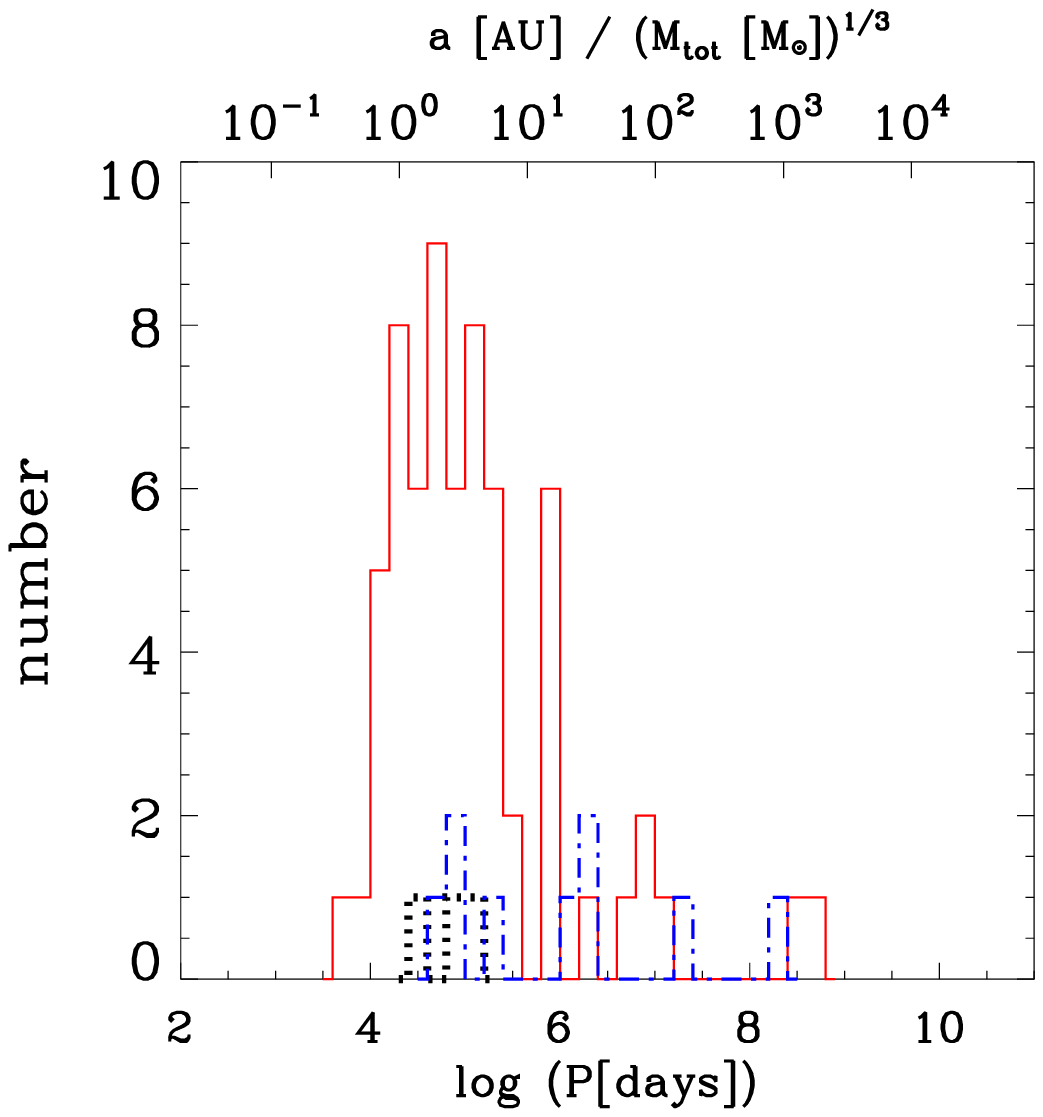}
 \caption{
{\it Left:}  Distribution of binary periods at 1000 yr in Minihalo 7.  Black dotted line represents the original minihalo, 
while red solid line represents the minihalo with increased resolution.  Blue dash-dotted line depicts the low-resolution case.
{\it Right:} Distribution of binary periods at 1000 yr in Minihalo 10.  Lines have the same meaning as in the left panel.
}
\label{P_histE}
\end{figure*}

The binary period distributions at $t_{\rm acc} = 1000$ yr are shown in Fig. \ref{P_histE}, while further comparisons between the original, low, and high-resolution cases are listed in Table \ref{tab1}.  The high-resolution case of Minihalo 7 resolves a greater number of binaries over a wider range of periods (red line in Fig. \ref{P_histE}), though in all three cases the median period is roughly $10^5 - 10^6$ days.  The high-resolution case of Minihalo 10 also has a wider range of periods, but the median period in both the original and high-resolution cases is $10^5$ days.  Lowering the resolution shifts the median period to $\sim 10^6$ days.

In the new minihalo test cases, 
some sinks satisfy the binary criterion (Section 2.5) with several different potential companion sinks.  Such a sink is counted as a member of more than one binary system, which is why the number of binary pairs $N_{\rm bin}$ is more than half of $N_{\rm frag}$, the total number of sinks remaining after 1000 yr (Table \ref{tab1}).  Instead of showing the binary fraction $f_{\rm bin}$ as defined in Equation 10, for each case we show the `companion fraction' $f_{\rm com}$, the fraction of sinks which have one or more companions.  In both minihaloes, regardless of resolution, $f_{\rm com}$ ranges from 50-100\%.  The percentage of sinks ejected from the system ranges from 0 to 50\%, but note that even in the single low-resolution case in which there were no ejections by 1000 yr, it is likely that  ejections will eventually occur at later times.  The high rate of ejections and binarity within Pop III clusters thus occurs not only over a range of minihaloes with a range of stellar accretion rates, but over a range of resolution scales as well.

\begin{table}
\centering
\begin{tabular}{cccccc}
\hline
Name &  $N_{\rm frag}$  &  $N_{\rm ejec}$ &  $N_{\rm bin}$  & $f_{\rm com}$ & $q_{\rm avg}$ \\
\hline
Minihalo 7-LR    &  3   &  0  &   2  & 100\%  & 0.57  \\
Minihalo 7       &  4   &  2  &   1  &  50\%  & 0.64  \\
Minihalo 7-HR    & 12   &  3  &  10  &  67\%  & 0.27   \\
Minihal 10-LR    &  9   &  2  &   9  &  78\%  & 0.32   \\
Minihal 10       &  7   &  2  &   3  &  71\%  & 0.26  \\
Minihal 10-HR    &   80   &  13   &   65   &    68\%   &  0.17     \\
\hline
\end{tabular}
\caption{Summary of binary statistics for low-resolution (LR), original, and high-resolution (HR) minihalos at $t_{\rm acc}$ = 1000 yr.}
\label{tab1}
\end{table}

\section{Caveats}
The results presented in the simulation have a number of caveats that should be noted.  First, these simulations followed only 5000 yr of protostellar mass accretion.  The distribution of sink masses and binary properties will continue to evolve beyond this point, as some of the sinks will continue to grow in mass and orbits may harden or undergo disruption through close encounters with other sinks.  We also cannot study binaries within the resolution limit of the simulation, 20 AU.  Thus, the sink mergers discussed previously may have instead become very tight and unresolved binaries.  Objects formed from subsequent fragmentation on sub-sink scales
also could not be addressed in this study.  
If we did presume that all sink mergers instead became tight binaries orbiting with semi-major axis of 10 AU, we would have a binary fraction of 73\%.  The average orbital radius of all binaries would then be 170 AU, and the average mass ratio would still be 0.34.

In addition, we do not include protostellar feedback in our simulations.  The H$_2$-dissociating and ionizing radiation emitted from the growing protostars may substantially change the characteristics of the stellar disk and eventually cut off mass flow onto the disk altogether (e.g. \citealt{hosokawaetal2011,smithetal2011,stacyetal2012}).
Another important physical process not included in our simulations was magnetic fields, which  may affect the angular momentum build-up of the protostellar disk, the subsequent fragmentation, and the mass flow onto the protostars (e.g. \citealt{tan&blackman2004,silk&langer2006, maki&susa2007,schleicheretal2010, federrathetal2011,suretal2012}; \nocite{schoberetal2012a}Schober et al. 2012a, \nocite{schoberetal2012b}2012b).  In the future we plan to enhance the physical realism of subseqent simulations by including these processes and resolving smaller scales.

\section{Summary and Conclusions}

We have examined the formation of Pop III multiple systems within ten different minihaloes extracted from a 1.4 Mpc (comoving) cosmological box.  
The properties of each stellar group varies between minihaloes, with mass functions ranging from flat to $\alpha=0.57$, and average stellar accretion rates varying by over a factor of three between minihaloes. 

Looking at the combined properties of the ten stellar systems, we find that 50\% of the protostars merge with larger protostars, while 50\% of the remaining protostars are ejected from their stellar disk. We furthermore find a binary fraction of $\sim$ 36\%, translating to a  53\% probability that any given star will have a companion.  Some of these binaries have semi-major axes that extend to 3000 AU.  The distribution of semi-major axes peaks slightly at $\sim$ 300 AU, while the mass ratio is nearly uniformly distributed between zero and one.
Recent work by Smith et al. (2012b)\nocite{smithetal2012b} implies that this result may be modified by the effects of heating and ionization from DM annihilation (DMA).  If their findings hold up to further investigation, DMA may suppress fragmentation and lead to wider binaries of $a>1000$ AU.  

Our simulation did not have sufficient resolution to resolve protostellar mergers or the formation of tight binaries,
as these processes occur on scales smaller than our resolution length of 20 AU.  If a significant fraction of these sink mergers did represent true protostellar mergers or the formation of contact binaries, this has important implications for the Pop III stellar evolution (see, e.g. \citealt{langer2012} and references therein).  
During a common-envelope phase, H-envelope removal by a close binary companion may allow for a star as low as 10 M$_{\odot}$ to become a WR star (e.g. \citealt{vanbeverenetal1998,crowther2007}).  
A stellar merger may rejuvenate the mass-gaining star, making it appear younger than its true age (e.g. \citealt{hellings1983,braun&langer1995, dray&tout2007}).  Alternatively, a star with an undermassive helium core may result (\citealt{braun&langer1995}).
Such stars may avoid the red supergiant phase and instead explode as blue supergiants.
Mass accretion through Roche-lobe overflow may also spin up a star, allowing for rotationally-induced mixing and possibly chemically homogeneous evolution, thereby altering the star's luminosity, temperature, and metal production (e.g. \citealt{ekstrometal2008,yoonetal2012}).   
In particular, rotationally induced mixing may alter the nucleosynthesis within a low-metallicity or primordial massive star to yield greater 
CNO-element abundances,
especially $^{14}$N (e.g. \citealt{ekstrometal2008,yoonetal2012}).  This additionally leads to a neutron excess in the core, enhancing the production of $s$-process elements (e.g. \citealt{pignatarietal2008}).

The binarity of Pop III stars also provides an important pathway for the formation of GRBs and high-mass X-ray binaries.  If one of the binary members is in the mass range to collapse into a BH, then a close companion may remove its hydrogen envelope and allow for penetration of a GRB jet to the surface (e.g. \citealt{leeetal2002,izzardetal2004}).  Once the BH remnant is left behind, the Roche lobe overflow from the companion enables the BH to sustain a high luminosity and X-ray flux. These sources could significantly contribute to reionization, and exert feedback effects during the assembly of the first galaxies (\citealt{mirabeletal2011,jeonetal2012}).

Pop III binarity is thus a crucial aspect to understanding Pop III evolution and feedback.  A significant percentage of Pop III stars exist in binaries and undergo stellar mergers.  Greater computational power will allow for future simulations which include more physical processes such as feedback and magnetic field growth, and which will also follow the evolution of Pop III multiple systems with increased resolution over longer periods of time.  Combined with observational constraints, this will yield an increasingly refined understanding of Pop III stars and their role in the early Universe.

\section*{Acknowledgements}
The authors wish to thank John Mather for insightful discussion during the development of this work.
AS is grateful for support from the JWST Postdoctoral Fellowship through the NASA Postdoctoral Program (NPP).  
VB acknowledges support from
NASA through Astrophysics Theory and Fundamental
Physics Program grant NNX09AJ33G and from NSF through
grant AST-1009928.
Resources supporting this work were provided by the NASA High-End Computing (HEC) Program through the NASA Advanced Supercomputing (NAS) Division at Ames Research Center.

\bibliographystyle{mn2e}
\bibliography{binaries}{}

\label{lastpage}

\end{document}